\documentclass[11pt,a4paper]{article}
\usepackage{amsmath,amssymb,graphicx,epsfig,cite,color,hyperref}
\usepackage[left=2.7cm,right=2.7cm,top=2.5cm,bottom=2.5cm]{geometry}
\usepackage{subfigure}
\usepackage{xcolor}

\usepackage{graphicx}
\usepackage{lscape}
\usepackage{appendix}
\usepackage{multirow}

\begin{document}

\def\ds{\displaystyle}
\def\beq{\begin{equation}}
\def\eeq{\end{equation}}
\def\bea{\begin{eqnarray}}
\def\eea{\end{eqnarray}}
\def\beeq{\begin{eqnarray}}
\def\eeeq{\end{eqnarray}}

\def\rar{\rightarrow} 
\def\nnb{\nonumber}

\def\ds{\displaystyle}
\def\beq{\begin{equation}}
\def\eeq{\end{equation}}
\def\bea{\begin{eqnarray}}
\def\eea{\end{eqnarray}}
\def\beeq{\begin{eqnarray}}
\def\eeeq{\end{eqnarray}}
\def\ve{\vert}
\def\vel{\left|}
\def\ver{\right|}
\def\nnb{\nonumber}
\def\ga{\left(}
\def\dr{\right)}
\def\aga{\left\{}
\def\adr{\right\}}
\def\lla{\left<}
\def\rra{\right>}
\def\rar{\rightarrow}
\def\lrar{\leftrightarrow}
\def\nnb{\nonumber}
\def\la{\langle}
\def\ra{\rangle}
\def\ba{\begin{array}}
\def\ea{\end{array}}
\def\tr{\mbox{Tr}}
\def\ssp{{\Sigma^{*+}}}
\def\sso{{\Sigma^{*0}}}
\def\ssm{{\Sigma^{*-}}}
\def\xis0{{\Xi^{*0}}}
\def\xism{{\Xi^{*-}}}
\def\qs{\la \bar s s \ra}
\def\qu{\la \bar u u \ra}
\def\qd{\la \bar d d \ra}
\def\qq{\la \bar q q \ra}
\def\gGgG{\la g^2 G^2 \ra}
\def\q{\gamma_5 \not\!q}
\def\x{\gamma_5 \not\!x}
\def\g5{\gamma_5}
\def\sb{S_Q^{cf}}
\def\sd{S_d^{be}}
\def\su{S_u^{ad}}
\def\sbp{{S}_Q^{'cf}}
\def\sdp{{S}_d^{'be}}
\def\sup{{S}_u^{'ad}}
\def\ssp{{S}_s^{'??}}

\def\sig{\sigma_{\mu \nu} \gamma_5 p^\mu q^\nu}
\def\fo{f_0(\frac{s_0}{M^2})}
\def\ffi{f_1(\frac{s_0}{M^2})}
\def\fii{f_2(\frac{s_0}{M^2})}
\def\O{{\cal O}}
\def\sl{{\Sigma^0 \Lambda}}
\def\es{\!\!\! &=& \!\!\!}
\def\ap{\!\!\! &\approx& \!\!\!}
\def\ar{&+& \!\!\!}
\def\ek{&-& \!\!\!}
\def\kek{\!\!\!&-& \!\!\!}
\def\cp{&\times& \!\!\!}
\def\se{\!\!\! &\simeq& \!\!\!}
\def\eqv{&\equiv& \!\!\!}
\def\kpm{&\pm& \!\!\!}
\def\kmp{&\mp& \!\!\!}
\def\mcdot{\!\cdot\!}
\def\erar{&\rightarrow&}



\renewcommand{\textfraction}{0.2}    
\renewcommand{\topfraction}{0.8}   

\renewcommand{\bottomfraction}{0.4}   
\renewcommand{\floatpagefraction}{0.8}
\newcommand\mysection{\setcounter{equation}{0}\section}

\def\baeq{\begin{appeq}}     \def\eaeq{\end{appeq}}  
\def\baeeq{\begin{appeeq}}   \def\eaeeq{\end{appeeq}}
\newenvironment{appeq}{\beq}{\eeq}   
\newenvironment{appeeq}{\beeq}{\eeeq}
\def\bAPP#1#2{
 \markright{APPENDIX #1}
 \addcontentsline{toc}{section}{Appendix #1: #2}
 \medskip
 \medskip
 \begin{center}      {\bf\LARGE Appendix #1 :}{\quad\Large\bf #2}
\end{center}
 \renewcommand{\thesection}{#1.\arabic{section}}
\setcounter{equation}{0}
        \renewcommand{\thehran}{#1.\arabic{hran}}
\renewenvironment{appeq}
  {  \renewcommand{\theequation}{#1.\arabic{equation}}
     \beq
  }{\eeq}
\renewenvironment{appeeq}
  {  \renewcommand{\theequation}{#1.\arabic{equation}}
     \beeq
  }{\eeeq}
\nopagebreak \noindent}

\def\eAPP{\renewcommand{\thehran}{\thesection.\arabic{hran}}}

\renewcommand{\theequation}{\arabic{equation}}
\newcounter{hran}
\renewcommand{\thehran}{\thesection.\arabic{hran}}

\def\bmini{\setcounter{hran}{\value{equation}}
\refstepcounter{hran}\setcounter{equation}{0}
\renewcommand{\theequation}{\thehran\alph{equation}}\begin{eqnarray}}
\def\bminiG#1{\setcounter{hran}{\value{equation}}
\refstepcounter{hran}\setcounter{equation}{-1}
\renewcommand{\theequation}{\thehran\alph{equation}}
\refstepcounter{equation}\label{#1}\begin{eqnarray}}


\newskip\humongous \humongous=0pt plus 1000pt minus 1000pt
\def\caja{\mathsurround=0pt}
 

\title{
         {\Large
                 {\bf
Beautiful mathematics for beauty-full and other multi-heavy hadronic systems 
                 }
         }
      }

\author{\vspace{1cm}\\
{\small
K. Azizi$^{1,2}$ \thanks {e-mail:
kazizi@dogus.edu.tr}~\,\,, 
A. R. Olamaei$^3$ \thanks {e-mail:
olamaei@jahromu.ac.ir}
\,\,, S. Rostami$^4$ \thanks
{e-mail: Asalrostami.phy@gmail.com}} \\
{\small $^1$ Department of Physics, Do\v{g}u\c{s} University, Acibadem-Kadik\"{o}y, 34722
Istanbul, Turkey }\\
{\small$^2$  School of Physics, Institute for Research in Fundamental 
Sciences (IPM)}\\
{\small P.~O.~Box 19395-5531, Tehran, Iran}\\
{\small $^3$ Department of Physics, Jahrom University, Jahrom, P.~ O.~ Box 74137-66171, Iran }\\
{\small $^4$ Young Researchers and Elites Club, South Tehran Branch, Islamic Azad University, Tehran, Iran}} 

\date{}

\begin{titlepage}
\maketitle
\thispagestyle{empty}

\begin{abstract}
In  most  non-perturbative methods in hadron 
physics the calculations are started with a correlation 
function in terms of some interpolating and transition currents in $ x $-space. 
For simplicity, the calculations are then transformed to 
the momentum space by a Fourier transformation.
To suppress the contributions of the higher 
states and continuum;
and enhance the ground state contribution,
 Borel transformation 
as well as continuum subtraction
are applied by the help of quark-hadron
duality assumption. 
In the present study we work out the
 mathematics required for these processes 
 in the case of light and multi-heavy 
 hadrons. We address a well-known problem 
 in subtraction of the effects of the higher states 
 and continuum and discuss how we find finite results without
 any divergence by using an appropriate representation
 of the modified Bessel functions, appearing 
 in the heavy quark propagator, and successive applications of the 
 Borel transformations,
 which lead to more suppression of the higher states
 and continuum contributions.
 The results obtained can be used in determination 
 of the spectroscopic and decay properties of the multi-heavy 
 standard and non-conventional (exotic) systems in many 
 non-perturbative methods, specially the QCD sum rules.
\end{abstract}

\end{titlepage}
\section{Introduction}
The traditional mesons and baryons as 
strong bound states of quarks and gluons are 
represented as the standard $ q\bar{q} $ and $ qqq $ states in 
terms of the valence quarks. There have been made  good progresses 
on identification and determination of many spectroscopic 
and decay properties of theses standard hadrons 
both experimentally and theoretically (for more information see for instance \cite{Aliev:2018hre,Aliev:2018ube,Azizi:2018dtb,Agaev:2017lip} and references therein).
Indeed, many ground and higher states at different 
light and heavy channels have been discovered in the experiment.
Roughly, all the light and heavy ground states mesons predicted 
by the quark model, have been observed.
Besides, all the light and single charmed ground state 
baryons together with some excited states have been 
detected by different experiments, as well. 
In the case of heavy $ b $-baryons, except the 
$ \Omega_b^* $ baryon with spin-$ \frac{3}{2} $, 
all single heavy baryons have been seen. 
For the standard baryons with two or three 
heavy quarks, however, only the $ \Xi_{cc} $ has been discovered.
For the first time, its existence was reported by SELEX collaboration 
\cite{Mattson:2002vu,Ocherashvili:2004hi}, 
but it was not been later confirmed by other groups. 
Recently, the LHCb
collaboration has  reported the observation of doubly 
heavy baryon $ \Xi^{++}_{cc} $
via the decay mode $ \Lambda_c^+ K^- \pi^+\pi^−$
with mass $ 3621.40\pm 0.72 (\text{stat.})\pm 0.27 (\text{syst.})\pm 0.14 (\Lambda_c^+)~\text{MeV}/c^2 $ \cite{Aaij:2017ueg}. 

Neither the quark model, nor the QCD as the theory 
of the strong interaction do not exclude the existence 
of the strong bound states out of the traditional 
$ q\bar{q} $ and $ qqq $ systems
\cite{Jaffe:1976ig,Jaffe:1976ih,Isgur:1984bm,DeRujula:1976zlg,Jaffe:2004ph}. 
The existence of such non-conventional (exotic) 
states were predicted by Jaffe \cite{Jaffe:1975fd,Jaffe:1976ig}.
Although they had been predicted more than forty 
years ago, the experimental searches had ended 
up in null results up to 2003, when the Belle 
Collaboration discovered the famous $ X(3872) $  
in the $ \pi^+\pi^-j/\psi $ invariant mass distribution from  
$ B^\pm \rightarrow K^\pm\pi^+\pi^-J/\psi $ 
decays  \cite{Choi:2003ue}.
This discovery had stimulated the experimental 
and theoretical investigations of the exotic states 
such that a plenty of  tetraquarks were discovered 
by different experimental collaborations.

Motivated by these progresses, in 2015, the LHCb 
Collaboration announced the observation of two 
pentaquark states $ P_c^{+}(4380) $ and $ P_c^{+} (4450) $ 
in the $ J/ \psi p$ spectrum of $ \Lambda_b\rightarrow J/ \psi ~p ~K^-  $ 
process \cite{Aaij:2015tga}. After discovery of the exotic states, 
many theoretical and experimental efforts have been devoted to 
the determination of the internal structure of these new objects. 
Despite a lot of studies, unfortunately, the nature and structure 
of most of exotic states remain unclear.
The studies predict a hidden charm/bottom structure together 
with two or three light quarks for the tetraquark or pentaquark 
states newly discovered by
experiments. However, there are many contradictory suggestions on
the quark organizations of 
these states in the literature (for more information see for example \cite{Azizi:2018bdv,Agaev:2017cfz,Agaev:2016srl} and references therein). Hence, we need more study on the spectroscopic 
and decay properties of the newly founded exotic and heavy baryonic states.

 Up to now, studies of hadrons with light and one heavy quark have well covered (see for
 instance Refs.~\cite{Aliev:2009ei,Aliev:2010dw,Aliev:2010yx,Aliev:2011ufa,Agaev:2016srl,Agaev:2017ywp,Roberts:2007ni,Valcarce:2008dr,Agaev:2016mjb,Lu:2016zhe,
 Wang:2010fq,Wang:2009cr,Agaev:2017jyt,Ebert:2011kk} and references therein) 
 but investigations of hadrons with two or more heavy flavour quarks are still
 rare (see for example Refs.~\cite{Aliev:2012iv,Aliev:2012nn,Aliev:2012ru,Aliev:2012tt,Aliev:2014lxa,Ebert:2002ig,Wei:2015gsa,Shi:2017dto,Wang:2017qvg} and references therein). 
In order to study the properties of 
heavy systems containing more heavy quarks
using some non-perturbative methods, especially
the QCD sum rules, we should develop
 the mathematics of heavy and light systems
 with more quarks. 
 The situation in calculations of the spectroscopic 
  parameters of such states and corresponding mathematics is better.
  Indeed, in \cite{Aliev:2012iv,Aliev:2012nn,Aliev:2012ru,Aliev:2012tt,Aliev:2014lxa}, the mathematics required for performing the
  Fourier and Borel transformations as well as the continuum
  subtraction related to the spectroscopic parameters of the doubly and triply heavy baryons has been well worked out. 
We present the  mathematics required to calculate all physical properties 
of the baryons with two-five heavy quarks,
especially their electromagnetic and strong decays
and their interactions with other particles.
A special case of these baryons with full $ b $ quarks
and without light quarks are called beauty-full
states (see for instance \cite{Anwar:2017toa}). We use the phrase `` beautiful mathematics'' in the title
of the manuscript to insist that the results are 
finite and the calculations do not include any divergence. 

  
Similar  calculations for the light and  single heavy  systems 
had been done 
for interaction with photon  ($ q^2=0 $) in Ref.~\cite{Ozpineci} and 
in the case of single heavy baryons with $ q^2\neq0 $ in Ref.~\cite{Aliev:2011ufa}.
We extend these calculations to  study the  heavy systems containing two-five
heavy quarks for all values of the transferred momentum squared.
For reader we repeat the calculations done in Refs.~\cite{Ozpineci,Aliev:2011ufa}
for the light and heavy systems to give an idea
about the calculations for simple systems.

This article is structured in the following way. 
Section \ref{LH} is
devoted to developing
the mathematics of the light and heavy systems,  each 
of them is explained in details in separate sub-sections.
In section \ref{concl} we present the concluding remarks.

\section{Mathematics for light and heavy systems}\label{LH} 

As we previously mentioned, in many non-perturbative methods, especially the
QCD sum rules (traditional SVZ and light cone QCD \cite{Shifman:1978bx,Balitsky:1989ry}),
in order to calculate the physical parameters of hadrons as strong
bound states of quarks and gluons, we start with a basic object called  correlation
function.
This function is expressed in terms of time ordering
product of some interpolating or transition
currents in  coordinate space. As an example let us consider the strong interaction of doubly heavy spin-1/2 baryons with 
light pseudoscalar mesons. The light-cone correlation function responsible for such vertices can be written as
\begin{eqnarray}\label{equ1} 
\Pi= i \int d^4x e^{ipx} \left< {\cal P}(q) \vert {\cal T} \left\{
\eta (x) \bar{\eta} (0) \right\} \vert 0 \right>~,
\end{eqnarray}
where $ {\cal P}(q) $ denotes the pseudoscalar mesons of momentum $ q $. In light-cone QCD sum rules we use the distribution amplitudes (DAs) of these mesons expanded in terms of wave functions having different twists (for details about DAs of the pseudoscalar particles please see the Appendix). In Eq. (\ref{equ1}) $ \eta $ represents the interpolating currents of the doubly heavy baryons. It is given as 
\begin{eqnarray}
 \eta^{S}&=&\frac{1}{\sqrt{2}}\epsilon_{abc}\Bigg\{(Q^{aT}Cq^b)\gamma_{5}Q'^c+
(Q'^{aT}Cq^b)\gamma_{5}Q^c+\beta(Q^{aT}C\gamma_{5}q^b)Q'^c+\beta(Q'^{aT}C
\gamma_{5}q^b)Q^c\Bigg\},\nonumber\\
\eta^{A}&=&\frac{1}{\sqrt{6}}\epsilon_{abc}\Bigg\{2(Q^{aT}CQ'^b)\gamma_{5}q^c+
(Q^{aT}Cq^b)\gamma_{5}Q'^c-(Q'^{aT}Cq^b)\gamma_{5}Q^c+2\beta(Q^{aT}C
\gamma_{5}Q'^b)q^c\nonumber\\
&+&\beta(Q^{aT}C\gamma_{5}q^b)Q'^c-\beta(Q'^{aT}C\gamma_{5}q^b)Q^c\Bigg\},
\end{eqnarray}
where $ S $ and $ A $ respectively represent the symmetric and anti-symmetric parts, $\beta$ is a mixing parameter with  $\beta=-1$ corresponding to the famous Ioffe current, $C$ is charge conjugation operator; and
$a$, $b$, and $c$ are the color indices. Here $Q/Q'$ and $q$ correspond to the heavy and
light quarks fields, respectively. The interpolating current with the  $u$ or $d$ quark
corresponds to the $\Xi_{QQq}$, but with $s$ indicates the $\Omega_{QQq}$ baryons, respectively.
Note  that in the symmetric part, both heavy
quarks can be identical or different, but in the anti-symmetric part two
heavy quarks should be different.
After insertion of the above currents into the correlation function and contracting
out the quark-pairs using the Wick's theorem, we get a result in terms
of the heavy and light quarks' propagators. For instance for the symmetric part we get
\begin{eqnarray}\label{tree expresion.m}
\Pi^{S}&=&i{\cal{A}} \epsilon_{abc}\epsilon_{a'b'c'}\int d^4x e^{i p x}\langle{\cal P}(q)\mid\Big\{-\gamma_{5}
S^{cb'}_{Q}S'^{ba'}_{q}S^{ac'}_{Q'}\gamma_{5}
-\gamma_{5}S^{cb'}_{Q'}S'^{ba'}_{q}S^{ac'}_{Q}\gamma_{5}
\nonumber\\&+&\gamma_{5}S^{cc'}_{Q'}\gamma_{5}Tr\Big[S^{ab'}_{Q}S'^{ba'}_{q}\Big]
+\gamma_{5}S^{cc'}_{Q}\gamma_{5}Tr\Big[S^{ab'}_{Q'}S'^{ba'}_{q}\Big]
\nonumber\\&+&\beta\Big( -\gamma_{5}S^{cb'}_{Q}\gamma_{5}S'^{ba'}_{q}S^{ac'}_{Q'}
-\gamma_{5}S^{cb'}_{Q'}\gamma_{5}S'^{ba'}_{q}S^{ac'}_{Q}-S^{cb'}_{Q}S'^{ba'}_{q}
\gamma_{5}S^{ac'}_{Q'}\gamma_{5}\nonumber\\&-&S^{cb'}_{Q'}S'^{ba'}_{q}\gamma_{5}
S^{ac'}_{Q}\gamma_{5}
+\gamma_{5}S^{cc'}_{Q'}Tr\Big[S^{ab'}_{Q}\gamma_{5}S'^{ba'}_{q}\Big]+S^{cc'}_{Q'}
\gamma_{5}Tr\Big[S^{ab'}_{Q}S'^{ba'}_{q}\gamma_{5}\Big]\nonumber\\&+&
\gamma_{5}S^{cc'}_{Q}Tr\Big[S^{ab'}_{Q'}\gamma_{5}S'^{ba'}_{q}\Big]+S^{cc'}_{Q}
\gamma_{5}Tr\Big[S^{ab'}_{Q'}S'^{ba'}_{q}\gamma_{5}\Big]\Big)\nonumber\\&+&
\beta^2\Big( -S^{cb'}_{Q}\gamma_{5}S'^{ba'}_{q}\gamma_{5}S^{ac'}_{Q'}-S^{cb'}_{Q'}
\gamma_{5}S'^{ba'}_{q}\gamma_{5}S^{ac'}_{Q}\nonumber\\&+&
S^{cc'}_{Q'}Tr\Big[S^{ba'}_{q}\gamma_{5}S'^{ab'}_{Q}\gamma_{5}\Big]+S^{cc'}_{Q}
Tr\Big[S^{ba'}_{q}\gamma_{5}S'^{ab'}_{Q'}\gamma_{5}\Big]
\Big)
\Big\}\mid 0\rangle,
\end{eqnarray}
where $S'=CS^TC$, with $ S $ being the heavy or light quark propagator, and $ \cal{A} $ is normalization constant. 
To proceed, we need to know the explicit expressions of the light and heavy quark propagators in $  x$-space. They are given as \cite{Balitsky:1987bk,Reinders:1984sr,Agaev:2016srl,Lee:2002jb} 
\begin{eqnarray}\label{prolight} 
S_q(x) \!\!\! &=& \!\!\! {i \rlap/x\over 2\pi^2 x^4} - {m_q\over 4 \pi^2 x^2} -
{\langle \bar q q \rangle \over 12} \left(1 - i {m_q\over 4} \rlap/x \right) -
{x^2\over 192} m_0^2 \langle \bar q q \rangle  \left( 1 -
i {m_q\over 6}\rlap/x \right) \nonumber \\
&&  - i g_s \int_0^1 du \left[{\rlap/x\over 16 \pi^2 x^2} G_{\mu \nu} (ux)
\sigma_{\mu \nu} - {i\over 4 \pi^2 x^2} u x^\mu G_{\mu \nu} (ux) \gamma^\nu
\right. \nonumber \\
&& \left.
 - i {m_q\over 32 \pi^2} G_{\mu \nu}(ux) \sigma^{\mu
 \nu} \left( \ln \left( {-x^2 \Lambda^2\over 4} \right) +
 2 \gamma_E \right) \right]~,
\end{eqnarray}
and
\begin{eqnarray}\label{proheavy} 
S_Q(x) \!\!\! &=& \!\!\! {m_Q^2 \over 4 \pi^2} {K_1(m_Q\sqrt{-x^2}) \over \sqrt{-x^2}} -
i {m_Q^2 \rlap/{x} \over 4 \pi^2 x^2} K_2(m_Q\sqrt{-x^2})\nonumber \\
&&-
ig_s \int {d^4k \over (2\pi)^4} e^{-ikx} \int_0^1
du \Bigg[ {\rlap/k+m_Q \over 2 (m_Q^2-k^2)^2} G^{\mu\nu} (ux)
\sigma_{\mu\nu}\nonumber \\
&& +
{u \over m_Q^2-k^2} x_\mu G^{\mu\nu} (ux)\gamma_\nu \Bigg]~,
\end{eqnarray}
where $\gamma_E $ is the Euler constant, $ G_{\mu \nu} $ is the gluon field strength tensor,  $\Lambda$ is the
scale parameter and $K_{\nu}$ in the heavy propagator denote the Bessel functions of the second kind.
By inserting the explicit expressions of the heavy and light propagators we end up with an expression
in   coordinate space that we need to transfer
it to the momentum space in order to make
the calculations easy and ready for the application of the
Borel transformation as well as the continuum subtraction
to enhance the ground state pole contribution and suppress
the effects of the higher state and continuum.
In the case of heavy systems with
more than one heavy quarks, using the integral
representation of the Bessel functions of the second kind,
appearing in the heavy quark propagator in the 
form
 \begin{equation}
\frac{K_{\nu }\left( m_{Q}\sqrt{-x^{2}}\right) }{\left( \sqrt{-x^{2}}\right)
^{\upsilon }}=\frac{1}{2}\int_{0}^{\infty }\frac{dt}{t^{\nu +1}}\exp \left[ -\frac{m_{Q}}{2}\left( t-\frac{x^{2}}{t}\right) \right],
\end{equation}
in Minkowski space with $ m_Q $
being the heavy quark mass, leads to a well-known problem and we end up with indeterminate results in the
calculations of the continuum subtraction 
(for details see the Appendix C of the Ref.~\cite{Ozdem:2017jqh}).
By choosing an appropriate representation of the Bessel functions
and applying successive Borel transformations
with the aim of more suppressing the unwanted contributions,
we show how we obtain a 
finite result for systems of multi-heavy quarks.
In the following, we present the mathematics required
for these processes in the light, single heavy, doubly heavy, triple
heavy, four heavy and five heavy systems. Note that we will use the free parts of the quark propagators to present the calculations, however,
the applied method is also valid when we include the interacting parts of the quark propagators. The interacting parts of the heavy quark propagator can also be written in terms of the modified Bessel functions \cite{Agaev:2016srl}.

 \subsection{Light systems}
In the following, we study the mathematics required for investigation of
 the light  hadrons containing the light ($ u,d,s $) quarks
 and their interaction with the photon and other particles.
Inserting  the light quarks' propagators
into the expression obtained after contracting out the quark pairs
in the correlation functions in many non-perturbative methods,
we get the following generic expression in  $ x $-space:
\begin{equation}\label{a1}
T_0 (p,q)= \int_0^1 du~ \int d^4x ~e^{iP.x} f(u) \frac{1}{(-x^2)^n},
\end{equation}
where $ P=p+uq $ with $ uq $ coming from
distribution amplitudes of the on-shell state, $ u $ is the momentum
fraction and $f(u) $ is a general function.
In the following for simplicity we omit the dependence of $ T_0 $ function on $ p$ and $q $.
We perform a Wick rotation   to go the Euclidean space
\begin{equation}\label{a4}
T_0= -i \int_0^1 ~du\int d^4x ~e^{-iP.x}~   f(u) \frac{1}{(x^2)^n}.
\end{equation}

Now we use the Schwinger parametrization
\begin{equation}\label{a5}
\frac{1}{A^n}=\frac{1}{\Gamma(n)}\int_0^\infty dt~t^{n-1} e^{-t A}~~~~~~~~A>0,
\end{equation}
and get
\begin{eqnarray}\label{a6}
T_0&=&\frac{-i}{\Gamma(n)}  \int_0^1 du  \int d^4x ~ f(u) \int_0^\infty dt~e^{-iP.x} e^{-tx^2}t^{n-1}.
\end{eqnarray}

The next step is to make the power of the exponential full-squared and perform 
the resultant Gaussian integral over four $ x $ using
\begin{equation}\label{a7}
\int d^4x~e^{-iP.x} e^{-tx^2} =(\frac{\pi}{t})^2 e^{\frac{-P^2}{4t}}.
\end{equation}


This leads to
\begin{equation}\label{a10}
T_0=\frac{-i}{\Gamma(n)}  \int_0^1 du~   f(u) \int_0^\infty dt~e^{\frac{-P^2}{4t}} (\frac{\pi}{t})^2t^{n-1}.
\end{equation}

Using the double Borel transformation with respect to $ (p+q)^2$ and $p^2 $
by the help of
\begin{equation}\label{a11}
{\cal B}_{M_1}{\cal B}_{M_2}e^{\frac{-P^2}{4t}}=\delta(\frac{1}{M_1^2}-\frac{u}{4t})\delta(\frac{1}{M_2^2}-\frac{\bar{u}}{4t})e^{\frac{q^2}{M_1^2+M_2^2}},
\end{equation}
and performing the $ u $ and $ t $ integration, one obtains
\begin{equation}\label{a12}
{\cal B}_{M_1}{\cal B}_{M_2}T_0= \frac{ -i 4^{2-n}\pi^2}{\Gamma(n)}
f(u_0) (M^2)^n  e^{\frac{q^2}{M_1^2+M_2^2}},
\end{equation}
where $ u_0=\frac{M_2^2}{M_1^2+M_2^2} $ and $M^2=\frac{M_1^2 M_2^2}{M_1^2+M_2^2} $.

\subsubsection*{\textit{Spectral representation and continuum subtraction for light systems}}

Our aim is to write the $ T_0 $ function in terms of a double dispersion integral
\begin{eqnarray}\label{a13}
T_0\left(  (p+q)^2,p^2\right)  = \int_0^\infty ds_1 \int_0^\infty ds_2 \frac{\rho_0 (s_1,s_2)}{[s_1-(p+q)^2][s_2-p^2]},
\end{eqnarray}
and find the spectral density $ \rho_0 (s_1,s_2) $. The final goal is the application of continuum 
subtraction in order to more suppress the contributions
of higher states and continuum.
The double Borel transformed form of the
$ T_0 $ function is written as
\begin{eqnarray}\label{a14}
T_0(M_1^2,M_2^2)  = \int_0^\infty ds_1 \int_0^\infty ds_2 \rho_0 (s_1,s_2)e^{-\frac{s_1}{M_1^2}}e^{-\frac{s_2}{M_2^2}}.
\end{eqnarray}
Now, let us discuss how  contributions of the continuum and  higher states are subtracted and the spectral  density $ \rho_0 (s_1,s_2) $ is obtained. To this end, we consider a generic term of the
 form 
\baeeq \label{a14}
\Pi_0=(M^2)^n f(u_0).
\eaeeq
The first step is to expand $f(u_0)$ as
\baeeq \label{a15}
f(u_0)=\sum_k a_ku_0^k.
\eaeeq

As a result we get
\baeeq \label{a16}
\Pi_0=\Bigg(\frac{M_1^2 M_2^2}{M_1^2+M_2^2}\Bigg)^n\sum_k a_k\Bigg(\frac{ M_2^2}{M_1^2+M_2^2}\Bigg)^k.
\eaeeq

Introducing new variables, $\sigma_i=\frac{1}{M_i^2}$ , we have
\baeeq \label{a17}
\Pi_0&=&\sum_k a_k \frac{\sigma_1^k}{(\sigma_1+\sigma_2)^{n+k}}\nonumber\\ 
&=&\sum_k  a_k \frac{\sigma_1^k}{\Gamma(n+k)}\int_0^\infty dt~
e^{-t(\sigma_1+\sigma_2)}t^{n+k-1}.
\eaeeq

Applying double Borel transformation 
with respect to $\sigma_1\rar\frac{1}{\tau_1}$ and $\sigma_2\rar\frac{1}{\tau_2}$
, we obtain the  spectral density as
\baeeq \label{a18}
{\cal B}_{1/\tau_1}{\cal B}_{1/\tau_2}\Pi_0&=&\sum_k a_k \frac{(-1)^k}{\Gamma(n+k)}\int_0^\infty dt~ t^{n+k-1}\Bigg((\frac{d}{dt})^k\delta(\tau_1-t)\Bigg)
\delta(\tau_1-t).
\eaeeq

We define the spectral density 
$ \rho_0 (s_1,s_2)={\cal B}_{1/\tau_1}{\cal B}_{1/\tau_2}\Pi_0$ 
with $\tau_1\rar s_1$ and $\tau_2\rar s_2$. Hence,
\baeeq \label{a19}
\rho_0(s_1,s_2)&=&\sum_k a_k \frac{(-1)^k}{\Gamma(n+k)}\int_0^\infty dt ~t^{n+k-1}\Bigg((\frac{d}{dt})^k\delta(s_1-t)\Bigg)
\delta(s_2-t).
\eaeeq \label{a20}

Performing integration over $t$, finally, we obtain the following expression for the double spectral density:
\baeeq
\rho_0(s_1,s_2)&=&\sum_k a_k \frac{(-1)^k}{\Gamma(n+k)} s_1^{n+k-1}\delta^k(s_2-s_1),
\eaeeq
where $ \delta^k(x)=\frac{d^k}{dx^k}\delta(x) $.
Using this spectral density, the continuum subtracted correlation 
function in the Borel scheme corresponding to the considered term can be written as:
\baeeq \label{a21}
\Pi_0^{sub}(M_1^2,M_2^2) =\int_0^{s_1} ds_1\int_0^{s_2}ds_2~\rho(s_1,s_2)e^{-s_1/M_1^2}e^{-s_2/M_2^2},
\eaeeq
where ``sub'' stands for subtracted. Defining new variables, $s_1= s v$ and $s_2= s(1- v)$, we get
\eAPP
\baeeq  \label{a22}
\Pi_0^{sub}(M_1^2,M_2^2)=\int_0^{s_0}ds\int_0^1 dv~\rho(s_1,s_2)se^{-sv/M_1^2}e^{-s(1-v)/M_2^2}.
\eaeeq
Finally, using the expression for the spectral density, one can get
\baeeq \label{a23}
\Pi_0^{sub}(M_1^2,M_2^2)&=&\sum_k a_k \frac{(-1)^k}{\Gamma(n+k)}\int_0^{s_0}ds \frac{s^{n-1}}{2}(\frac{d}{dv})^k\Bigg[v^{n+k-1}e^{-sv/M_1^2}e^{-s(1-v)/M_2^2}\Bigg]_{v=1/2}.\nonumber\\
\eaeeq

\subsection{Systems with one heavy quark}
In the following, we study the hadrons containing a single heavy quark
with some light quarks.
Let us consider the following generic term:                                                                                                                                                                     
\begin{equation}\label{b1}
T_1= \int_0^1 du \int d^4x ~e^{iP.x}~   f(u) \frac{K_\nu(m_Q\sqrt{-x^2})}{(\sqrt{-x^2})^n},
\end{equation}
where $ m_Q $ is the mass of heavy quark and $ K_\nu $ is the modified
 Bessel function of the second kind in order $ \nu $.
 The $ K_\nu $ function comes from the $ x $-representation
 of the heavy quark propagator.
  Using the following integral representation
 of the modified Bessel function 
\begin{equation}\label{b2}
K_\nu(m_Q\sqrt{-x^2})=\frac{\Gamma(\nu+ 1/2)~2^\nu}{\sqrt{\pi}m_Q^\nu}\int_0^\infty dt~\cos(m_Qt)\frac{(\sqrt{-x^2})^\nu}{(t^2-x^2)^{\nu+1/2}},
\end{equation}
 we have
\begin{equation}\label{b3}
T_1=  \frac{\Gamma(\nu+ 1/2)~2^\nu}{\sqrt{\pi}m_Q^\nu}\int_0^1 du\int d^4x~e^{iP.x}~f(u)\int_0^\infty dt~ \frac{\cos(m_Qt)}{(t^2-x^2)^{\nu+1/2} (x^2)^{\frac{n-\nu}{2}}}.
\end{equation}

We perform a Wick rotation   to go to the Euclidean space
\begin{equation}\label{b4}
T_1= \frac{\Gamma(\nu+ 1/2)~2^\nu}{\sqrt{\pi}m_Q^\nu}\int_0^1 du\int d^4 x~(-i)e^{-iP.x}~f(u)\int_0^\infty dt~ \frac{\cos(m_Qt)}{(t^2+x^2)^{\nu+1/2} (x^2)^{\frac{n-\nu}{2}}}.
\end{equation}

Using the identify
\begin{equation} \label{b5}
\frac{1}{A^n}=\frac{1}{\Gamma(n)}\int_0^\infty dt~t^{n-1} e^{-t A}~~~~~~~~A>0,
\end{equation}
Eq. (\ref{b4}) can be reformulated  as
\begin{eqnarray}\label{b6}
T_1&=& \frac{2^\nu (-i)}{\Gamma(\frac{n-\nu}{2})\sqrt{\pi}m_Q^\nu}\int_0^1 du\int d^4x~e^{-iP.x}~f(u)\nonumber\\
&\times&\int_0^\infty dt~ \int_0^\infty dy \int_0^\infty dv~v^{\nu-\frac{1}{2
}} e^{-v (x^2+t^2)}y^{\frac{n-\nu}{2}-1} e^{-y (x^2)}\cos(m_Qt).\nonumber\\
\end{eqnarray}

In the next step, we performe the Gaussian integral over four-$ x $ and $ t $
\begin{equation}\label{b7}
\int d^4x~e^{-iP.x} e^{-vx^2} e^{-yx^2}=(\frac{\pi}{y+v})^2e^{\frac{-P^2}{4(y+v)}},
\end{equation}

and
\begin{equation}\label{b9}
\int_0^\infty dt~\cos({mt}) ~e^{-vt^2} =\frac{e^{\frac{-m^2}{4v}}\sqrt{\pi}}{2\sqrt{v}}.
\end{equation}

As a result, we get:
\begin{equation}\label{b10}
T_1= \frac{2^{\nu-1}\pi^2 (-i)}{\Gamma(\frac{n-\nu}{2})m_Q^\nu}
\int_0^1 du~f(u)\int_0^\infty  dy \int_0^\infty dv~\frac{e^{\frac{-m^2}{4v}}v^{\nu-1} y^{\frac{n-\nu}{2}-1} e^{\frac{-P^2}{4(y+v)}}}{(v+y)^2}.
\end{equation}

We  introduce the new variables $ \rho $ and $ \lambda $, defined by

\begin{equation} \label{b11}
~~~~~~~~~~~~~~~~~~~~~~~~~\rho=v+y,~~~~~~~~~~~~~\lambda=\frac{y}{v+y},
\end{equation}
which leads to
\begin{equation}\label{b12}
T_1=\frac{2^{\nu-1}\pi^2 (-i)}{\Gamma(\frac{n-\nu}{2})m_Q^\nu}
\int_0^1 du~f(u)\int_0^\infty  d\rho\int_0^1 d\lambda~e^{\frac{-m^2}{4\rho(1-\lambda)}}\rho^{\frac{n+\nu}{2}-3}(1-\lambda)^{\nu-1}  \lambda^{\frac{n-\nu}{2}-1} e^{\frac{-P^2}{4\rho}}.
\end{equation}

Using the Borel transformation of the exponential
\begin{equation} \label{b13}
{\cal B}_{M_1}{\cal B}_{M_2}e^{\frac{-P^2}{4\rho}}=\delta(\frac{1}{M_1^2}-\frac{u}{4\rho})\delta(\frac{1}{M_2^2}-\frac{\bar{u}}{4\rho})e^{\frac{q^2}{M_1^2+M_2^2}},
\end{equation}
and performing the $ u $ and $ \rho $ integrals, one obtain
\begin{eqnarray}\label{b14}
{\cal B}_{M_1}{\cal B}_{M_2}T_1&=&-i  \frac{2^{2-n}\pi^2}{m_Q^\nu}
f(u_0)(M^2)^{\frac{\nu+n}{2}} \int_0^1 d\lambda~\frac{e^{\frac{-m^2}{M^2(1-\lambda)}}(1-\lambda)^{\nu-1} \lambda^{\frac{n-\nu}{2}-1}}{\Gamma(\frac{n-\nu}{2})} e^{\frac{q^2}{M_1^2+M_2^2}},
\end{eqnarray}
where $ u_0=\frac{M_2^2}{M_1^2+M_2^2} $ and $M^2=\frac{M_1^2 M_2^2}{M_1^2+M_2^2} $.  By replacing $\lambda=x^2$, we get
\begin{eqnarray}\label{b14}
{\cal B}_{M_1}{\cal B}_{M_2}T_1&=&-i  \frac{2^{2-n}\pi^2}{m_Q^\nu}
f(u_0)(M^2)^{\frac{\nu+n}{2}}e^{\frac{q^2}{M_1^2+M_2^2}}
 \int_0^1 dx~\frac{e^{\frac{-m^2}{M^2(1-x^2)}}(1-x^2)^{\nu-1} x^{n-\nu-1}}{\Gamma(\frac{n-\nu}{2})}.
\end{eqnarray}

In the last step, we change the variable $ \frac{1}{1-x^2}\rightarrow t $ and we get
\begin{equation}\label{b16}
{\cal B}_{M_1}{\cal B}_{M_2}T_1=-i  \frac{2^{2-n}\pi^2}{m_Q^\nu}
f(u_0)(M^2)^{\frac{\nu+n}{2}} e^{\frac{q^2}{M_1^2+M_2^2}} \psi(\alpha,\beta,\frac{m_Q^2}{M^2}),
\end{equation}
where
\begin{equation} \label{b171}
\psi(\alpha,\beta,\frac{m_Q^2}{M^2})=\frac{1}{\Gamma(\alpha)}\int_1^\infty dt e^{-t\frac{m_Q^2}{M^2}} t^{\beta-\alpha-1}(t-1)^{\alpha-1},
\end{equation}
with $\alpha=\frac{n-\nu}{2}$ and $\beta=1-\nu$.

\subsection*{\textit{Spectral representation and continuum subtraction  for  systems containing one heavy quark }}

Now, let us discuss how  contribution of the continuum and  
higher states are subtracted. 
As is obvious from Eq.~(\ref{b16}),
the generic term has the
 form 
\baeeq \label{b18}
\Pi_1={\cal C}(M^2)^{\frac{\nu+n}{2}} f(u_0) \psi(\alpha,\beta,\frac{m_Q^2}{M^2}),
\eaeeq
where
\baeeq 
{\cal C}=-i  \frac{2^{2-n}\pi^2}{m_Q^\nu}.
\eaeeq

The first step again is to expand $f(u_0)$ as
\baeeq \label{b19}
f(u_0)=\sum_k a_ku_0^k.
\eaeeq

As a result we get
\baeeq \label{b20}
\Pi_1={\cal C}\Bigg(\frac{M_1^2 M_2^2}{M_1^2+M_2^2}\Bigg)^{\frac{\nu+n}{2}}\sum_k a_k\Bigg(\frac{ M_2^2}{M_1^2+M_2^2}\Bigg)^k\frac{1}{\Gamma(\alpha)}\int_1^\infty dt
e^{-t\frac{m_Q^2}{M^2}} t^{\beta-\alpha-1}(t-1)^{\alpha-1}.
\eaeeq \label{b21}

Introducing new variables, $\sigma_i=\frac{1}{M_i^2}$ , we have
\baeeq \label{b22}
\Pi_1&=&{\cal C}\sum_k a_k \frac{\sigma_1^k}{(\sigma_1+\sigma_2)^{\frac{\nu+n}{2}+k}}\frac{1}{\Gamma(\alpha)}\int_1^\infty dt ~
e^{-t m_Q^2(\sigma_1+\sigma_2)} t^{\beta-\alpha-1}(t-1)^{\alpha-1}\nonumber\\
&=&{\cal C}\sum_k a_k \frac{\sigma_1^k}{\Gamma(\frac{\nu+n}{2}+k)\Gamma(\alpha)}\int_1^\infty dt~
 e^{-t m_Q^2(\sigma_1+\sigma_2)}t^{\beta-\alpha-1}(t-1)^{\alpha-1}\int_0^\infty dl e^{-l(\sigma_1+\sigma_2)}l^{\frac{\nu+n}{2}+k-1}\nonumber\\
&=&{\cal C}\sum_k a_k \frac{\sigma_1^k}{\Gamma(\frac{\nu+n}{2}+k)\Gamma(\alpha)}\int_1^\infty dt~ 
 t^{\beta-\alpha-1}(t-1)^{\alpha-1}\int_0^\infty dl l^{\frac{\nu+n}{2}+k-1}e^{-(l+t m_Q^2)(\sigma_1+\sigma_2)}\nonumber\\
&=&{\cal C}\sum_k a_k \frac{(-1)^k}{\Gamma(\frac{\nu+n}{2}+k)\Gamma(\alpha)}\int_1^\infty dt~ 
 t^{\beta-\alpha-1}(t-1)^{\alpha-1}\nonumber\\
&\times&
 \int_0^\infty dl~l^{\frac{\nu+n}{2}+k-1}\Bigg((\frac{d}{dl})^k~e^{-(l+t m_Q^2)\sigma_1}\Bigg)e^{-(l+t m_Q^2)\sigma_2}.\nonumber\\
\eaeeq

Applying double Borel transformation 
with respect to $\sigma_1\rar\frac{1}{\tau_1}$ and $\sigma_2\rar\frac{1}{\tau_2}$, we obtain 
\baeeq \label{b23}
{\cal B}_{1/\tau_1}{\cal B}_{1/\tau_2}\Pi_1&=&{\cal C}\sum_k a_k \frac{(-1)^k}{\Gamma(\frac{\nu+n}{2}+k)\Gamma(\alpha)}\int_1^\infty dt~ 
 t^{\beta-\alpha-1}(t-1)^{\alpha-1}\int_0^\infty dl~ l^{\frac{\nu+n}{2}+k-1}\nonumber\\
&\times& \Bigg((\frac{d}{dl})^k\delta(\tau_1-(l+t m_Q^2))\Bigg)\delta(\tau_2-(l+t m_Q^2)).
\eaeeq

The double spectral density $ \rho_1(s_1,s_2) $ is found by replacing 
$\tau_1\rar s_1$ and $\tau_2\rar s_2$ in the expression as
\baeeq \label{b24}
\rho_1(s_1,s_2)&=&{\cal C}\sum_k a_k \frac{(-1)^k}{\Gamma(\frac{\nu+n}{2}+k)\Gamma(\alpha)}\int_1^\infty dt~ 
 t^{\beta-\alpha-1}(t-1)^{\alpha-1}\int_0^\infty dl~l^{\frac{\nu+n}{2}+k-1}\nonumber\\
&\times&\Bigg((\frac{d}{dl})^k\delta(s_1-(l+t m_Q^2))\Bigg)\delta(s_2-(l+t m_Q^2)).
\eaeeq

Performing the integration over $l$, finally we obtain the following expression for the double spectral density:
\baeeq \label{b25}
\rho_1(s_1,s_2)&=&{\cal C}\sum_k a_k \frac{(-1)^k}{\Gamma(\frac{\nu+n}{2}+k)\Gamma(\alpha)}\int_1^\infty dt~ 
 t^{\beta-\alpha-1}(t-1)^{\alpha-1}(s_1-t m_Q^2)^{\frac{\nu+n}{2}+k-1}\nonumber\\
&\times&\Bigg((\frac{d}{ds_1})^k\delta(s_2-s_1)\Bigg)\theta(s_1-t m_Q^2),
\eaeeq
or
\baeeq \label{b26}
\rho_1(s_1,s_2)&=&{\cal C}\sum_k a_k \frac{(-1)^k}{\Gamma(\frac{\nu+n}{2}+k)\Gamma(\alpha)}\int_1^{s_1/m_Q^2} dt~ 
 t^{\beta-\alpha-1}(t-1)^{\alpha-1}(s_1-t m_Q^2)^{\frac{\nu+n}{2}+k-1}\nonumber\\
&\times&\Bigg((\frac{d}{ds_1})^k\delta(s_2-s_1)\Bigg).
\eaeeq

Using this spectral density, the continuum subtracted correlation function in the Borel scheme corresponding to the considered term can be written as:
\baeeq \label{b27}
\Pi^{sub}(M_1^2,M_2^2)=\int_{m_Q^2}^{s_0}ds_1\int_{m_Q^2}^{s_0}ds_2~\rho(s_1,s_2)e^{-s_1/M_1^2}e^{-s_2/M_2^2}.
\eaeeq

Defining new variables, $s_1=2 s v$ and $s_2=2 s(1- v)$, we get
\baeeq \label{b28}
\Pi^{sub}(M_1^2,M_2^2)=\int_{m_Q^2}^{s_0}ds\int_0^1 dv~\rho(s_1,s_2)(4s)e^{-2sv/M_1^2}e^{-2s(1-v)/M_2^2}.
\eaeeq

Using the expression for the spectral density, one can get
\baeeq \label{b29}
\Pi_1^{sub}(M_1^2,M_2^2)&=&{\cal C}\sum_k a_k \frac{(-1)^k}{\Gamma(\frac{\nu+n}{2}+k)\Gamma(\alpha)}\int_{m_Q^2}^{s_0}ds\int_0^1 dv\frac{1}{2^ks^k}\Bigg((\frac{d}{dv})^k\delta(v-1/2)\Bigg)\nonumber\\
&\times&\int_1^{2sv/m_Q^2} dt ~
 t^{\beta-\alpha-1}(t-1)^{\alpha-1}(2sv-t m_Q^2)^{\frac{\nu+n}{2}+k-1}e^{-2sv/M_1^2}e^{-2s(1-v)/M_2^2}.\nonumber\\
\eaeeq

By integrating by part over $v$, finally we obtain
\baeeq \label{b30}
\Pi_1^{sub}(M_1^2,M_2^2)&=&-i  \sum_k a_k \frac{2^{2-n}\pi^2}{m_Q^\nu \Gamma(\frac{\nu+n}{2}+k)\Gamma(\alpha)}\int_{m_Q^2}^{s_0}ds\frac{1}{2^ks^k}\Bigg[(\frac{d}{dv})^k\int_1^{2sv/m_Q^2} dt ~
 t^{\beta-\alpha-1}(t-1)^{\alpha-1}\nonumber\\
&\times&(2sv-t m_Q^2)^{\frac{\nu+n}{2}+k-1}e^{-2sv/M_1^2}e^{-2s(1-v)/M_2^2}\Bigg]_{v=1/2}.\nonumber\\
\eaeeq

\subsection{Systems containing two heavy quarks}
In the following, we study hadrons containing two heavy quarks and some light
quarks.
Let us consider again the following generic term:                                                                                                                                                                     
\begin{equation}\label{c1}
T_2= \int_0^1 du\int d^4x ~e^{iP.x}~   A(u) \frac{K_\nu(m_{1Q}\sqrt{-x^2})~K_\mu(m_{2Q}\sqrt{-x^2})}{(\sqrt{-x^2})^n},
\end{equation}
where $ m_{iQ} $'s are the masses of heavy quarks and $ K_\nu $ and $ K_\mu $
are the modified
 Bessel functions of order $ \nu $ and $ \mu $, respectively. Using the integral representation
 of the modified Bessel functions from Eq.~(\ref{b2}),
 we have
\begin{eqnarray}\label{c3}
T_2&=&\frac{\Gamma(\nu+ 1/2)\Gamma(\mu+ 1/2)~2^{\mu+\nu}}{\pi m_{1Q}^\nu m_{2Q}^\mu}\int_0^1 du\int d^4x~e^{iP.x}~A(u)\nonumber\\
&\times&\int_0^\infty dt_2 \int_0^\infty dt_1~ \frac{\cos(m_{1Q}t_1)\cos(m_{2Q}t_2)}{(t_1^2-x^2)^{\nu+1/2}(t_2^2-x^2)^{\mu+1/2}(x^2)^{\frac{n-\nu-\mu}{2}}}.
\end{eqnarray}

We again perform a Wick rotation   to go the Euclidean space
\begin{eqnarray}\label{c4}
T_2&=&(-i)\frac{\Gamma(\nu+ 1/2)\Gamma(\mu+ 1/2)~2^{\mu+\nu}}{\pi m_{1Q}^\nu m_{2Q}^\mu}\int_0^1 du\int d^4x~e^{-iP.x}~A(u)\nonumber\\
&\times&\int_0^\infty dt_2\int_0^\infty dt_1~ \frac{\cos(m_{1Q}t_1)\cos(m_{2Q}t_2)}{(t_1^2+x^2)^{\nu+1/2}(t_2^2+x^2)^{\mu+1/2}(x^2)^{\frac{n-\nu-\mu}{2}}}.
\end{eqnarray}

Using the Schwinger integral representation presented in Eq.~(\ref{b5}),
Eq.~(\ref{c4}) can be reformulated as 
\begin{eqnarray}\label{c6}
T_2&=&(-i) \frac{2^{\mu+\nu}}{\pi m_{1Q}^\nu m_{2Q}^\mu \Gamma(\frac{n-\nu-\mu}{2}) }\int_0^1 du\int d^4x~e^{-iP.x}~A(u) \int_0^\infty dt_2\int_0^\infty dt_1 \cos(m_{1Q}t_1)\cos(m_{2Q}t_2)\nonumber\\
&\times& \int_0^\infty dy_2 \int_0^\infty dy_1
\int_0^\infty dy_0~y_0^{(\frac{n-\nu-\mu}{2})-1} e^{-y_0 x^2}~y_1^{\nu-\frac{1}{2}} e^{-y_1 (x^2+t_1^2)}y_2^{\mu-\frac{1}{2}} e^{-y_2 (x^2+t_2^2)}.\nonumber\\
\end{eqnarray}


Performing the Gaussian integral over four-$ x $, $ t_1 $ and $ t_2 $, we get
\begin{eqnarray}\label{c10}
T_2&=&(-i) \frac{2^{\mu+\nu-2} \pi^2}{m_{1Q}^\nu m_{2Q}^\mu \Gamma(\frac{n-\nu-\mu}{2}) }\int_0^1 du~A(u)  \int_0^\infty dy_0~y_0^{(\frac{n-\nu-\mu}{2})-1}\nonumber\\
&\times& \int_0^\infty dy_2 \int_0^\infty dy_1~\frac{y_1^{\nu-1} e^{-\frac{m^2_{1Q}}{4y_1}}y_2^{\mu-1} e^{-\frac{m^2_{2Q}}{4y_2}}e^{\frac{-P^2}{4(y_0+y_1+y_2)}}}{(y_0+y_1+y_2)^2}.\nonumber\\
\end{eqnarray}

Now we  introduce the variables $ \rho $, $ v $ and $ u $, defined by

\begin{equation} \label{c11}
~~~~~~~~~~~\rho=y_0+y_1+y_2,~~~~~~~~v=\frac{y_1}{y_0+y_1+y_2},~~~~~~~~w=\frac{y_2}{y_0+y_1+y_2},
\end{equation}
which leads to

\begin{eqnarray}\label{c12}
T_2&=&(-i) \frac{2^{\mu+\nu-2}\pi^2}{ m_{1Q}^\nu m_{2Q}^\mu \Gamma(\frac{n-\nu-\mu}{2}) }\int_0^1 du~A(u)  \int_0^1 dw~(\rho(1-v-w))^{(\frac{n-\nu-\mu}{2})-1}\nonumber\\
&\times& \int_0^\infty d\rho \int_0^1 dv~(\rho v)^{\nu-1} e^{-\frac{m^2_{1Q}}{4\rho v}}(\rho w)^{\mu-1} e^{-\frac{m^2_{2Q}}{4\rho w}}e^{\frac{-P^2}{4\rho}}.\nonumber\\
\end{eqnarray}

Applying  again the double Borel transformations
with respect to $(p+q)^2\rightarrow M_1^2 $ and 
$ p^2\rightarrow M_2^2 $ and performing the
integrals over the variables $ u $ and $ \rho $,
using the resultant Dirac Deltas, we obtain

\begin{eqnarray}\label{c14}
{\cal B}_{M_1}{\cal B}_{M_2}T_2&=&(-i) \frac{2^{2-n} \pi^2}{ m_{1Q}^\nu m_{2Q}^\mu\Gamma(\frac{n-\nu-\mu}{2}) }A(u_0)(M^2)^{\frac{n+\nu+\mu}{2}} e^{\frac{q^2}{M_1^2+M_2^2}}
\int_0^1 dw \int_0^1 dv~\nonumber\\
&\times &(1-v-w)^{(\frac{n-\nu-\mu}{2})-1}e^{-\frac{m^2_{1Q}}{M^2 v}} v^{\nu-1} e^{-\frac{m^2_{2Q}}{M^2 w}}
(w)^{\mu-1} .
\end{eqnarray}

The new variables 
\begin{equation} \label{c15}
~~~~~~~~~~~~~~~v=z(1-y),~~~~~~~~~~~~~~~~w=zy,
\end{equation}
change the above expression to
\begin{eqnarray}\label{c16}
{\cal B}_{M_1}{\cal B}_{M_2}T_2&=&(-i) \frac{2^{2-n} \pi^2}{ m_{1Q}^\nu m_{2Q}^\mu\Gamma(\frac{n-\nu-\mu}{2}) }A(u_0)(M^2)^{\frac{n+\nu+\mu}{2}} e^{\frac{q^2}{M_1^2+M_2^2}}
\int_0^2  dz \int_0^1  dy~\nonumber\\
&\times& z^{\nu+\mu -1} (1- z)^{(\frac{n-\nu-\mu}{2})-1} (1-y)^{\nu-1} y^{\mu -1} e^{-\frac{m^2_{1Q}}{M^2 z(1- y)}}  e^{-\frac{m^2_{2Q}}{M^2  zy}}.
\end{eqnarray}

\subsection*{\textit{Spectral representation and continuum subtraction for  systems containing two heavy quarks}}

Now, let us discuss how  contributions of the  higher states and continuum are 
subtracted. We consider again a generic term of the
 form 
\baeeq \label{c19}
\Pi_2&=&{\cal C}(M^2)^{\frac{n+\nu+\mu}{2}} A(u_0)\int_0^2  dz \int_0^1  dy~z^{\nu+\mu -1} (1- z)^{(\frac{n-\nu-\mu}{2})-1}\nonumber\\
&\times&  (1-y)^{\nu-1} y^{\mu -1} e^{-\frac{m^2_{1Q}}{M^2 z(1- y)}}  e^{-\frac{m^2_{2Q}}{M^2  zy}}.
\eaeeq
where
\baeeq \label{c20}
{\cal C}=(-i) \frac{2^{2-n} \pi^2}{ m_{1Q}^\nu m_{2Q}^\mu\Gamma(\frac{n-\nu-\mu}{2}) }.
\eaeeq

 The first step is to expand $A(u_0)$ as
\baeeq \label{c20}
A(u_0)=\sum_k a_ku_0^k.
\eaeeq
As a result we get
\baeeq \label{c21}
\Pi_2&=&{\cal C}\Bigg(\frac{M_1^2 M_2^2}{M_1^2+M_2^2}\Bigg)^{\frac{n+\nu+\mu}{2}}\sum_k
 a_k\Bigg(\frac{ M_2^2}{M_1^2+M_2^2}\Bigg)^k\int_0^2 dz\int_0^1  dy~\nonumber\\
&\times& z^{\nu+\mu -1} (1- z)^{(\frac{n-\nu-\mu}{2})-1} (1-y)^{\nu-1} y^{\mu -1} e^{-\frac{m^2_{1Q}}{M^2 z(1- y)}}  e^{-\frac{m^2_{2Q}}{M^2  zy}}.
\eaeeq

Introducing new variables, $\sigma_i=\frac{1}{M_i^2}$ , we have
\baeeq \label{c22}
\Pi_2&=&{\cal C}\sum_k a_k \frac{\sigma_1^k}{(\sigma_1+\sigma_2)^{\frac{n+\nu+\mu}{2}+k}}\int_0^2 dz\int_0^1  dy~ z^{\nu+\mu -1} (1- z)^{(\frac{n-\nu-\mu}{2})-1}\nonumber\\
&\times& (1-y)^{\nu-1} y^{\mu -1} e^{-\frac{m^2_{1Q}(\sigma_1+\sigma_2)}{ z(1- y)}}  e^{-\frac{m^2_{2Q}(\sigma_1+\sigma_2)}{  zy}}\nonumber\\
&=&{\cal C}\sum_k a_k \frac{\sigma_1^k}{\Gamma(\frac{n+\nu+\mu}{2}+k)}\int_0^2 dz \int_0^1  dy~ z^{\nu+\mu -1} (1- z)^{(\frac{n-\nu-\mu}{2})-1}\nonumber\\
&\times& (1-y)^{\nu-1} y^{\mu -1} e^{-\frac{m^2_{1Q}(\sigma_1+\sigma_2)}{ z(1- y)}}  e^{-\frac{m^2_{2Q}(\sigma_1+\sigma_2)}{  zy}}\int_0^\infty dl e^{-l(\sigma_1+\sigma_2)}l^{\frac{n+\nu+\mu}{2}+k-1}\nonumber\\
&=&{\cal C}\sum_k a_k \frac{\sigma_1^k}{\Gamma(\frac{n+\nu+\mu}{2}+k)}\int_0^2 dz \int_0^1  dy~ z^{\nu+\mu -1} (1- z)^{(\frac{n-\nu-\mu}{2})-1}\nonumber\\
&\times& (1-y)^{\nu-1} y^{\mu -1}
\int_0^\infty  dl ~l^{\frac{n+\nu+\mu}{2}+k-1}e^{-(l+\frac{m^2_{1Q}}{ z(1- y)}+\frac{m^2_{2Q}}{  zy})(\sigma_1+\sigma_2)}\nonumber\\
&=&{\cal C}\sum_k a_k \frac{(-1)^k}{\Gamma(\frac{n+\nu+\mu}{2}+k)}\int_0^2 dz  \int_0^1  dy~ z^{\nu+\mu -1} (1- z)^{(\frac{n-\nu-\mu}{2})-1}
\nonumber\\
&& (1-y)^{\nu-1} y^{\mu -1}\int_0^\infty dl~l^{\frac{n+\nu+\mu}{2}+k-1}\Bigg((\frac{d}{dl})^ke^{-(l+\frac{m^2_{1Q}}{z(1- y)}+\frac{m^2_{2Q}}{ z y})\sigma_1}\Bigg)e^{-(l+\frac{m^2_{1Q}}{z(1- y)}+\frac{m^2_{2Q}}{ z y})\sigma_2}.\nonumber\\
\eaeeq

Applying double Borel transformation with respect to $\sigma_1\rar\frac{1}{\tau_1}$ and $\sigma_2\rar\frac{1}{\tau_2}$, we obtain the  spectral density
\baeeq \label{c23}
 {\cal B}_{1/\tau_1}{\cal B}_{1/\tau_2}\Pi_2&=&{\cal C}\sum_k a_k \frac{(-1)^k}{\Gamma(\frac{n+\nu+\mu}{2}+k)}\int_0^2 dz \int_0^1  dy~ z^{\nu+\mu -1} \nonumber\\
&\times&(1- z)^{(\frac{n-\nu-\mu}{2})-1} (1-y)^{\nu-1} y^{\mu -1}\int_0^\infty dl l^{\frac{n+\nu+\mu}{2}+k-1}\nonumber\\
&\times&\Bigg((\frac{d}{dl})^k\delta(\tau_1-(l+\frac{m^2_{1Q}}{z(1- y)}+\frac{m^2_{2Q}}{ z y})\Bigg)\delta(\tau_2-(l+\frac{m^2_{1Q}}{z(1- y)}+\frac{m^2_{2Q}}{ z y})),\nonumber\\
\eaeeq
where the spectral density $ \rho_2(s_1,s_2) $ is found by
$\tau_1\rar s_1$ and $\tau_2\rar s_2$ in this relation
\baeeq \label{c24}
\rho_2(s_1,s_2)&=&{\cal C}\sum_k a_k \frac{(-1)^k}{\Gamma(\frac{n+\nu+\mu}{2}+k)}\int_0^2 dz \int_0^1  dy~ z^{\nu+\mu -1}\nonumber\\
&\times&  (1- z)^{(\frac{n-\nu-\mu}{2})-1}(1-y)^{\nu-1} y^{\mu -1}\int_0^\infty dl l^{\frac{n+\nu+\mu}{2}+k-1}\nonumber\\
&\times&\Bigg((\frac{d}{dl})^k\delta(s_1-(l+\frac{m^2_{1Q}}{z(1- y)}+\frac{m^2_{2Q}}{ z y})\Bigg)\delta(s_2-(l+\frac{m^2_{1Q}}{z(1- y)}+\frac{m^2_{2Q}}{ z y})).\nonumber\\
\eaeeq

Performing integration over $l$, finally we obtain the following expression for the double spectral density:
\baeeq \label{c25}
\rho_2(s_1,s_2)&=&{\cal C}\sum_k a_k \frac{(-1)^k}{\Gamma(\frac{n+\nu+\mu}{2}+k)}\int_0^2 dz \int_0^1  dy~ z^{\nu+\mu -1}
\nonumber\\
&\times& (1- z)^{(\frac{n-\nu-\mu}{2})-1} (1-y)^{\nu-1} y^{\mu -1}\Bigg(s_1-\frac{m^2_{1Q}}{z(1- y)}-\frac{m^2_{2Q}}{ z y}\Bigg)^{\frac{n+\nu+\mu}{2}+k-1}\nonumber\\
&\times&
\theta(s_1-\frac{m^2_{1Q}}{z(1- y)}-\frac{m^2_{2Q}}{ z y})
(\frac{d}{ds_1})^k\delta(s_2-s_1).\nonumber\\
\eaeeq

If we look at the integrand out of the Step function $ \theta $,
it diverges inside the  bound of $ z $.
However, this divergence is removed considering the step function 
coming from the successive application of the double Borel
transformations by modifying the limits of the integral over  $ z $
and removing the points leading to the divergences out of the boundaries.
Hence, we get  finite results for the values of 
$ n, k, \mu,$ and $ \nu $.
Using this spectral density, the continuum subtracted correlation function in the Borel scheme corresponding to the considered term can be written as
\baeeq \label{c26}
\Pi_2^{sub}(M_1^2,M_2^2)=\int_{s_L}^{s_0}ds_1\int_{s_L}^{s_0}ds_2~\rho(s_1,s_2)e^{-s_1/M_1^2}e^{-s_2/M_2^2},
\eaeeq
where $ s_L=(m_{1Q}+m_{2Q})^2$. Defining new variables, $s_1=2 s v$ and $s_2=2 s(1- v)$, we get
\baeeq \label{c27}
\Pi_2^{sub}(M_1^2,M_2^2)=\int_{s_L}^{s_0}ds\int_0^1 dv~\rho(s_1,s_2)(4s)e^{-2sv/M_1^2}e^{-2s(1-v)/M_2^2}.
\eaeeq

Using the expression for the spectral density, one can get
\baeeq \label{c28}
\Pi_2^{sub}(M_1^2,M_2^2)&=&{\cal C}\sum_k a_k \frac{(-1)^k}{\Gamma(\frac{n+\nu+\mu}{2}+k)}\int_{s_L}^{s_0}ds \int_0^1 dv\frac{1}{2^ks^k}\Bigg((\frac{d}{dv})^k\delta(v-1/2)\Bigg)\nonumber\\
&\times&\int_0^2 dz \int_0^1  dy~ z^{\nu+\mu -1} (1- z)^{(\frac{n-\nu-\mu}{2})-1} (1-y)^{\nu-1} y^{\mu -1}
\nonumber\\
&\times&\Bigg(2sv-\frac{m^2_{1Q}}{z(1- y)}-\frac{m^2_{2Q}}{ z y}\Bigg)^{\frac{n+\nu+\mu}{2}+k-1}~e^{-2sv/M_1^2}e^{-2s(1-v)/M_2^2}\nonumber\\
&\times&\theta(2sv-\frac{m^2_{1Q}}{z(1- y)}-\frac{m^2_{2Q}}{ z y}).
\eaeeq

Integrating over $v$, finally we obtain
\baeeq \label{c29}
\Pi_2^{sub}(M_1^2,M_2^2)&=&(-i) \frac{2^{2-n} \pi^2}{ m_{1Q}^\nu m_{2Q}^\mu\Gamma(\frac{n-\nu-\mu}{2}) }\sum_k a_k \frac{1}{\Gamma(\frac{n+\nu+\mu}{2}+k)}\nonumber\\
&\times& \int_{s_L}^{s_0}ds \frac{1}{2^ks^k} 
\Bigg[(\frac{d}{dv})^k    
\int_0^2 dz~\theta(2sv-\frac{m^2_{1Q}}{z(1- y)}-\frac{m^2_{2Q}}{ z y})\nonumber\\
&\times& \int_0^1  dy~ z^{\nu+\mu -1} (1- z)^{(\frac{n-\nu-\mu}{2})-1}
 (1-y)^{\nu-1} y^{\mu -1}\nonumber\\
&\times&(2sv-\frac{m^2_{1Q}}{z(1- y)}-\frac{m^2_{2Q}}{ z y})^{\frac{n+\nu+\mu}{2}+k-1}e^{-2sv/M_1^2}e^{-2s(1-v)/M_2^2}\Bigg]_{v=1/2}.
\eaeeq
\subsection{ Systems containing three heavy quarks}
In the following, we study hadrons containing triple heavy quarks
with some light quarks.
The general form of the correlation function in this case is
\begin{equation}\label{d1}
T_3=\int_0^1 du\int d^4x ~e^{iP.x} ~   A(u) \frac{K_\nu(m_{1Q}\sqrt{-x^2})~K_\mu(m_{2Q}\sqrt{-x^2})~K_\lambda(m_{3Q}\sqrt{-x^2})}{(\sqrt{-x^2})^n}.
\end{equation}

Using the definition of the modified Bessel function from Eq.~(\ref{b2}),
we obtain
\begin{eqnarray}\label{d3}
T_3&=&\frac{\Gamma(\nu+ 1/2)\Gamma(\mu+ 1/2)\Gamma(\lambda+ 1/2)~2^{\mu+\nu+\lambda}}{(\pi)^{3/2} m_{1Q}^\nu m_{2Q}^\mu m_{3Q}^\lambda}\int_0^1 du\int d^4x~e^{iP.x}~A(u)\nonumber\\
&\times &\int_0^\infty dt_3\int_0^\infty dt_2 \int_0^\infty dt_1~ \frac{\cos(m_{1Q}t_1)\cos(m_{2Q}t_2)\cos(m_{3Q}t_3)}{(t_1^2-x^2)^{\nu+1/2}(t_2^2-x^2)^{\mu+1/2}(t_3^2-x^2)^{\lambda+1/2}(-x^2)^{\frac{n-\nu-\mu-\lambda}{2}}},\nonumber\\
\end{eqnarray}
where a Wick rotation and usage of the  Schwinger integral
representation leads to
\begin{eqnarray}\label{d6}
T_3&=&(-i) \frac{2^{\mu+\nu+\lambda}}{(\pi)^{3/2} m_{1Q}^\nu m_{2Q}^\mu m_{3Q}^\lambda \Gamma(\frac{n-\nu-\mu-\lambda}{2})}\int_0^1 du\int d^4x~e^{-iP.x}~A(u)\int_0^\infty dt_3\int_0^\infty dt_2 \nonumber\\
&\times&\int_0^\infty dt_1 \int_0^\infty dy_0\int_0^\infty dy_3 \int_0^\infty dy_2 \int_0^\infty dy_1~y_1^{\nu-\frac{1}{2}} e^{-y_1 (x^2+t_1^2)}y_2^{\mu-\frac{1}{2}} e^{-y_2 (x^2+t_2^2)}\nonumber\\
&\times&y_3^{\lambda-\frac{1}{2}} e^{-y_3 (x^2+t_3^2)}y_0^{\frac{n-\nu-\mu-\lambda}{2}-1} e^{-y_0x^2}\cos(m_{1Q}t_1)\cos(m_{2Q}t_2)\cos(m_{3Q}t_3).\nonumber\\
\end{eqnarray}
 Performing the $ t_i $'s and $ x $-Gaussian integrals,
we get
\begin{eqnarray}\label{d10}
T_3&=&(-i) \frac{2^{\mu+\nu+\lambda-3}\pi^2}{ m_{1Q}^\nu m_{2Q}^\mu m_{3Q}^\lambda\Gamma(\frac{n-\nu-\mu-\lambda}{2})}\int_0^1 du~A(u) \int_0^\infty dy_0~y_0^{\frac{n-\nu-\mu-\lambda}{2}-1}\nonumber\\
&\times& \int_0^\infty dy_3 \int_0^\infty dy_2 \int_0^\infty dy_1~\frac{y_1^{\nu-1} e^{-\frac{m^2_{1Q}}{4y_1}}y_2^{\mu-1} e^{-\frac{m^2_{2Q}}{4y_2}} y_3^{\lambda-1} e^{-\frac{m^2_{3Q}}{4y_3}}e^{\frac{-P^2}{4(y_0+y_1+y_2+y_3)}}}{(y_0+y_1+y_2+y_3)^2}.\nonumber\\
\end{eqnarray}

We  introduce the variables $ \rho $, $v $,  $r $ and  $w $ defined by
\begin{eqnarray}\label{d110}
~~~\rho&=&y_0+y_1+y_2+y_3,~~v=\frac{y_1}{y_0+y_1+y_2+y_3},~~r=\frac{y_2}{y_0+y_1+y_2+y_3},
\nonumber\\~~w&=&\frac{y_3}{y_0+y_1+y_2+y_3},
\end{eqnarray}
which leads to
\begin{eqnarray}\label{d11}
T_3&=&(-i) \frac{2^{\mu+\nu+\lambda-3}\pi^2}{ m_{1Q}^\nu m_{2Q}^\mu m_{3Q}^\lambda\Gamma(\frac{n-\nu-\mu-\lambda}{2})}\int_0^1 du~A(u) \int_0^1 dw~ (\rho (1-v-r-w))^{\frac{n-\nu-\mu-\lambda}{2}-1}\nonumber\\
&\times& \int_0^\infty~\rho d\rho \int_0^1 dv\int_0^1 dr~(\rho v)^{\nu-1} e^{-\frac{m^2_{1Q}}{4\rho v}}(\rho r)^{\mu-1} e^{-\frac{m^2_{2Q}}{4\rho r}}(\rho w)^{\lambda-1} e^{-\frac{m^2_{3Q}}{4\rho w}}e^{\frac{-P^2}{4\rho}}.\nonumber\\
\end{eqnarray}

Now we apply  the Borel transformation of the exponential $ e^{\frac{-P^2}{4\rho}} $,
and perform the $ u $ and $ \rho $ integrals. As a result we get
\begin{eqnarray}\label{d13}
{\cal B}_{M_1}{\cal B}_{M_2}T_3&=&(-i) \frac{2^{1-n}\pi^2}{ m_{1Q}^\nu m_{2Q}^\mu m_{3Q}^\lambda \Gamma(\frac{n-\nu-\mu-\lambda}{2})}A(u_0)(M^2)^{\frac{n+\nu+\mu+\lambda}{2}}
e^{\frac{q^2}{M_1^2+M_2^2}}\int_0^1 dw\int_0^1 dv~\nonumber\\
 &\times&  \int_0^1 dr~ (1-v-r-w)^{\frac{n-\nu-\mu-\lambda}{2}-1} v^{\nu-1} e^{-\frac{m^2_{1Q}}{M^2 v}} r^{\mu-1} e^{-\frac{m^2_{2Q}}{M^2 r}} w^{\lambda-1} e^{-\frac{m^3_{3Q}}{M^2 w}}.\nonumber\\
\end{eqnarray}

The variables $ x $, $ y $ and $ z $, defined by
\begin{equation}\label{d14}
~~~~~~~~~~~~~~~v=zx(1-y),~~~~~~~~~~r=zxy,~~~~~~~~~~w=z(1-x),
\end{equation}
help us end up in the final form of the double Borel transformed  $ T $ function as
\begin{eqnarray}\label{d15}
{\cal B}_{M_1}{\cal B}_{M_2}T_3&=&(-i)  \frac{2^{1-n}\pi^2}{ m_{1Q}^\nu m_{2Q}^\mu m_{3Q}^\lambda \Gamma(\frac{n-\nu-\mu-\lambda}{2})}A(u_0)(M^2)^{\frac{n+\nu+\mu+\lambda}{2}}e^{\frac{q^2}{M_1^2+M_2^2}}\int_0^1 dx\int_0^1 dy~\nonumber\\
 &\times&  \int_0^3 dz(1-z)^{\frac{n-\nu-\mu-\lambda}{2}-1} z^{\nu+\mu+\lambda-1}x^{\nu+\mu-1}(1-x)^{\lambda-1}(1-y)^{\nu-1} y^{\mu-1}\nonumber\\
&\times& e^{-\frac{m^2_{1Q}}{M^2 zx(1-y)}}  e^{-\frac{m^2_{2Q}}{M^2 zxy}}
 e^{-\frac{m^2_{3Q}}{M^2 z(1-x)}}.
\end{eqnarray}

\subsection*{\textit{Spectral representation and continuum subtraction  for  systems containing three heavy quarks}}
To suppress the contribution of the higher states and  continuum
in this case we consider a general form as fallows
\baeeq \label{d16}
\Pi_3&=&{\cal C}(M^2)^{\frac{n+\nu+\mu+\lambda}{2}} A(u_0)\int_0^1 dx
\int_0^1 dy \int_0^3dz(1-z)^{\frac{n-\nu-\mu-\lambda}{2}-1} z^{\nu+\mu+\lambda-1}\nonumber\\
&\times& x^{\nu+\mu-1}(1-x)^{\lambda-1}(1-y)^{\nu-1} y^{\mu-1}e^{-\frac{m^2_{1Q}}{M^2 zx(1-y)}}  e^{-\frac{m^2_{2Q}}{M^2 zxy}}
 e^{-\frac{m^2_{3Q}}{M^2 z(1-x)}}.
\eaeeq 
where
\baeeq 
{\cal C}=(-i)  \frac{2^{1-n}\pi^2}{ m_{1Q}^\nu m_{2Q}^\mu m_{3Q}^\lambda \Gamma(\frac{n-\nu-\mu-\lambda}{2})}
\eaeeq 

Expressing $  A(u_0)$ function in series, leads to
\baeeq \label{d18}
\Pi_3&=&{\cal C}\Bigg(\frac{M_1^2 M_2^2}{M_1^2+M_2^2}\Bigg)^{\frac{n+\nu+\mu+\lambda}{2}}\sum_k a_k\Bigg(\frac{ M_2^2}{M_1^2+M_2^2}\Bigg)^k\int_0^1 dx
\int_0^1 dy \int_0^3dz(1-z)^{\frac{n-\nu-\mu-\lambda}{2}-1} \nonumber\\
&\times&z^{\nu+\mu+\lambda-1} x^{\nu+\mu-1}(1-x)^{\lambda-1}(1-y)^{\nu-1} y^{\mu-1}e^{-\frac{m^2_{1Q}}{M^2 zx(1-y)}}  e^{-\frac{m^2_{2Q}}{M^2 zxy}}
 e^{-\frac{m^2_{3Q}}{M^2 z(1-x)}}.
\eaeeq

We introduce the new variables, $\sigma_i=\frac{1}{M_i^2}$, 
and rearrange the terms
\baeeq \label{d19}
\Pi_3&=&{\cal C}\sum_k a_k \frac{\sigma_1^k}{(\sigma_1+\sigma_2)^{\frac{n+\nu+\mu+\lambda}{2}+k}}\int_0^1 dx
\int_0^1 dy \int_0^3dz(1-z)^{\frac{n-\nu-\mu-\lambda}{2}-1} z^{\nu+\mu+\lambda-1}\nonumber\\
&\times& x^{\nu+\mu-1}(1-x)^{\lambda-1}(1-y)^{\nu-1} y^{\mu-1}e^{-\frac{m^2_{1Q}(\sigma_1+\sigma_2)}{zx(1-y)}}  e^{-\frac{m^2_{2Q}(\sigma_1+\sigma_2)}{ zxy}}
 e^{-\frac{m^2_{3Q}(\sigma_1+\sigma_2)}{ z(1-x)}}\nonumber\\
&=&{\cal C}\sum_k a_k \frac{\sigma_1^k}{\Gamma(\frac{n+\nu+\mu+\lambda}{2}+k)}\int_0^1 dx
\int_0^1 dy \int_0^3dz(1-z)^{\frac{n-\nu-\mu-\lambda}{2}-1} z^{\nu+\mu+\lambda-1}\nonumber\\
&\times &x^{\nu+\mu-1}(1-x)^{\lambda-1}(1-y)^{\nu-1} y^{\mu-1}e^{-\frac{m^2_{1Q}(\sigma_1+\sigma_2)}{zx(1-y)}}  e^{-\frac{m^2_{2Q}(\sigma_1+\sigma_2)}{ zxy}}
 e^{-\frac{m^2_{3Q}(\sigma_1+\sigma_2)}{ z(1-x)}}
 \nonumber\\
&\times & \int_0^\infty dl e^{-l(\sigma_1+\sigma_2)}l^{\frac{n+\nu+\mu+\lambda}{2}+k-1}\nonumber\\
&=&{\cal C}\sum_k a_k \frac{\sigma_1^k}{\Gamma(\frac{n+\nu+\mu+\lambda}{2}+k)}\int_0^1 dx
\int_0^1 dy \int_0^3dz(1-z)^{\frac{n-\nu-\mu-\lambda}{2}-1} z^{\nu+\mu+\lambda-1}
x^{\nu+\mu-1}(1-x)^{\lambda-1}\nonumber\\
&\times& \int_0^\infty  dl l^{\frac{n+\nu+\mu+\lambda}{2}+k-1}e^{-(l+\frac{m^2_{1Q}}{zx(1-y)}+\frac{m^2_{2Q}}{ zxy}+\frac{m^2_{3Q}}{ z(1-x)})(\sigma_1+\sigma_2)}(1-y)^{\nu-1} y^{\mu-1}\nonumber\\
&=&{\cal C}\sum_k a_k \frac{(-1)^k}{\Gamma(\frac{n+\nu+\mu+\lambda}{2}+k)}\int_0^1 dx
\int_0^1 dy \int_0^3dz(1-z)^{\frac{n-\nu-\mu-\lambda}{2}-1} z^{\nu+\mu+\lambda-1}
x^{\nu+\mu-1}(1-x)^{\lambda-1}\nonumber\\
&\times &\int_0^\infty dl~l^{\frac{n+\nu+\mu+\lambda}{2}+k-1}\Bigg((\frac{d}{dl})^ke^{-(l+\frac{m^2_{1Q}}{zx(1-y)}+\frac{m^2_{2Q}}{ zxy}+\frac{m^2_{3Q}}{ z(1-x)})\sigma_1}\Bigg)\nonumber\\
&\times &e^{-(l+\frac{m^2_{1Q}}{zx(1-y)}+\frac{m^2_{2Q}}{ zxy}+\frac{m^2_{3Q}}{ z(1-x)})\sigma_2}(1-y)^{\nu-1} y^{\mu-1}.\nonumber\\
\eaeeq

Applying double Borel transformation with respect to
$\sigma_1\rar\frac{1}{\tau_1}$ and $\sigma_2\rar\frac{1}{\tau_2}$, 
gives us
\baeeq \label{d20}
{\cal B}_{1/\tau_1}{\cal B}_{1/\tau_2}\Pi_3&=&{\cal C}\sum_k a_k \frac{(-1)^k}{\Gamma(\frac{n+\nu+\mu+\lambda}{2}+k)}\int_0^1 dx
\int_0^1 dy \int_0^3dz(1-z)^{\frac{n-\nu-\mu-\lambda}{2}-1} z^{\nu+\mu+\lambda-1}
\nonumber\\
&\times &x^{\nu+\mu-1}(1-x)^{\lambda-1}(1-y)^{\nu-1} y^{\mu-1}
\int_0^\infty dl~l^{\frac{n+\nu+\mu+\lambda}{2}+k-1}\nonumber\\
&\times &\Bigg((\frac{d}{dl})^k\delta(\tau_1-(l+\frac{m^2_{1Q}}{zx(1-y)}+\frac{m^2_{2Q}}{ zxy}+\frac{m^2_{3Q}}{ z(1-x)})\Bigg)\nonumber\\
&\times &\delta\Big(\tau_2-(l+\frac{m^2_{1Q}}{zx(1-y)}+\frac{m^2_{2Q}}{ zxy}+\frac{m^2_{3Q}}{ z(1-x)})\Big),\nonumber\\
\eaeeq
where the spectral density $ \rho_3(s_1,s_2) $ is found
from the same manner as the previous sections as
\baeeq \label{d21}
\rho_3(s_1,s_2)&=&{\cal C}\sum_k a_k \frac{(-1)^k}{\Gamma(\frac{n+\nu+\mu+\lambda}{2}+k)}\int_0^1 dx
\int_0^1 dy \int_0^3dz(1-z)^{\frac{n-\nu-\mu-\lambda}{2}-1} z^{\nu+\mu+\lambda-1}
\nonumber\\
&\times &x^{\nu+\mu-1}(1-x)^{\lambda-1} (1-y)^{\nu-1} y^{\mu-1}
\int_0^\infty dl~l^{\frac{n+\nu+\mu+\lambda}{2}+k-1}\nonumber\\
&\times &\Bigg((\frac{d}{dl})^k\delta(s_1-(l+\frac{m^2_{1Q}}{zx(1-y)}+\frac{m^2_{2Q}}{ zxy}+\frac{m^2_{3Q}}{ z(1-x)})\Bigg)\nonumber\\
&\times &\delta\Big(s_2-(l+\frac{m^2_{1Q}}{zx(1-y)}+\frac{m^2_{2Q}}{ zxy}+\frac{m^2_{3Q}}{ z(1-x)})\Big).\nonumber\\
\eaeeq
Performing integration over $l$, finally we obtain the following expression for the double spectral density:
\baeeq \label{d22}
\rho_3(s_1,s_2)&=&{\cal C}\sum_k a_k \frac{(-1)^k}{\Gamma(\frac{n+\nu+\mu+\lambda}{2}+k)}\int_0^1 dx
\int_0^1 dy \int_0^3dz(1-z)^{\frac{n-\nu-\mu-\lambda}{2}-1} z^{\nu+\mu+\lambda-1}x^{\nu+\mu-1}
\nonumber\\
&\times &(1-x)^{\lambda-1}(1-y)^{\nu-1} y^{\mu-1}\Bigg(s_1-\frac{m^2_{1Q}}{zx(1-y)}-\frac{m^2_{2Q}}{ zxy}-\frac{m^2_{3Q}}{ z(1-x)}\Bigg)^{\frac{n+\nu+\mu+\lambda}{2}+k-1}\nonumber\\
&\times&\Bigg((\frac{d}{ds_1})^k\delta(s_2-s_1)\Bigg)\theta\Big(s_1-\frac{m^2_{1Q}}{zx(1-y)}-\frac{m^2_{2Q}}{ zxy}-\frac{m^2_{3Q}}{ z(1-x)}\Big).
\eaeeq
Using this spectral density, the continuum subtracted correlation function in the Borel scheme corresponding to the considered term can be written as:
\baeeq \label{d23}
\Pi_3^{sub}(M_1^2,M_2^2)=\int_{s_L}^{s_0}ds_1\int_{s_L}^{s_0}ds_2~\rho(s_1,s_2)e^{-s_1/M_1^2}e^{-s_2/M_2^2},
\eaeeq
where $ s_L=(m_{1Q}+m_{2Q}+m_{3Q})^2$.
Defining new variables, $s_1=2 s v$ and $s_2=2 s(1- v)$, we get
\eAPP
\baeeq \label{d24}
\Pi_3^{sub}(M_1^2,M_2^2)=\int_{s_L}^{s_0} ds\int_0^1 dv~\rho(s_1,s_2)(4s)e^{-2sv/M_1^2}e^{-2s(1-v)/M_2^2}.
\eaeeq
Using the expression for the spectral density, one can get
\baeeq \label{d25}
\Pi_3^{sub}(M_1^2,M_2^2)&=&{\cal C}\sum_k a_k \frac{(-1)^k}{\Gamma(\frac{n+\nu+\mu+\lambda}{2}+k)}\int_{s_L}^{s_0} ds\int_0^1 dv\frac{1}{2^ks^k}\Bigg((\frac{d}{dv})^k\delta(v-1/2)\Bigg)\nonumber\\
&\times&\int_0^1 dx
\int_0^1 dy \int_0^3dz(1-z)^{\frac{n-\nu-\mu-\lambda}{2}-1} z^{\nu+\mu+\lambda-1}
x^{\nu+\mu-1}(1-x)^{\lambda-1}\nonumber\\
&\times&(1-y)^{\nu-1} y^{\mu-1}\Bigg(2sv-\frac{m^2_{1Q}}{zx(1-y)}-\frac{m^2_{2Q}}{ zxy}-\frac{m^2_{3Q}}{ z(1-x)}\Bigg)^{\frac{n+\nu+\mu+\lambda}{2}+k-1}\nonumber\\
&\times&\theta\Big(2sv-\frac{m^2_{1Q}}{zx(1-y)}-\frac{m^2_{2Q}}{ zxy}-\frac{m^2_{3Q}}{ z(1-x)}\Big)
e^{-2sv/M_1^2}e^{-2s(1-v)/M_2^2}.
\eaeeq
Integrating over $v$, finally we obtain
\baeeq \label{d26}
\Pi_3^{sub}(M_1^2,M_2^2)&=&(-i) \sum_k a_k \frac{2^{1-n}\pi^2}{m_{1Q}^\nu m_{2Q}^\mu m_{3Q}^\lambda \Gamma(\frac{n-\nu-\mu-\lambda}{2})\Gamma(\frac{n+\nu+\mu+\lambda}{2}+k)}\nonumber\\
&\times&  \int ds\frac{1}{2^ks^k} 
\Bigg[(\frac{d}{dv})^k  e^{-2sv/M_1^2}e^{-2s(1-v)/M_2^2}
\nonumber\\
&\times&  \int_0^3 dz~\theta\Big(2sv-\frac{m^2_{1Q}}{zx(1-y)}-\frac{m^2_{2Q}}{ zxy}-\frac{m^2_{3Q}}{ z(1-x)}\Big)\nonumber\\
&\times&\int_0^1 dx
\int_0^1 dy (1-z)^{\frac{n-\nu-\mu-\lambda}{2}-1} z^{\nu+\mu+\lambda-1}
x^{\nu+\mu-1}(1-x)^{\lambda-1}
(1-y)^{\nu-1} y^{\mu-1}\nonumber\\
&\times&\Bigg(2sv-\frac{m^2_{1Q}}{zx(1-y)}-\frac{m^2_{2Q}}{ zxy}-\frac{m^2_{3Q}}{ z(1-x)}\Bigg)^{\frac{n+\nu+\mu+\lambda}{2}+k-1}\Bigg]_{v=1/2}.
\eaeeq

\subsection{Systems containing four heavy quarks}
In the following, we study hadrons containing four heavy 
and some light quarks.
We start with the generic term:                                                                                                                                                                     
\begin{equation}\label{e1}
T_4= \int_0^1 du\int d^4x ~e^{iP.x}~   A(u) \frac{K_\nu(m_{1Q}\sqrt{-x^2})~K_\mu(m_{2Q}\sqrt{-x^2})~K_\lambda(m_{3Q}\sqrt{-x^2})~K_\eta(m_{4Q}\sqrt{-x^2})}{(\sqrt{-x^2})^n},
\end{equation}
where $ m_{iQ} $ are again the masses of the heavy quarks. We
can rewrite Eq.~(\ref{e1}) as
\begin{eqnarray}\label{e3}
T_4&=&\frac{\Gamma(\nu+ 1/2)\Gamma(\mu+ 1/2)\Gamma(\lambda+ 1/2)\Gamma(\eta+ 1/2) ~2^{\mu+\nu+\lambda+\eta}}{(\pi)^2 m_{1Q}^\nu m_{2Q}^\mu m_{3Q}^\lambda m_{4Q}^\eta}\nonumber\\
&\times &\int_0^1 du\int d^4x~e^{iP.x}~A(u)\frac{1}{(-x^2)^\frac{n-\mu-\nu-\lambda-\eta}{2}}\int_0^\infty dt_4\int_0^\infty dt_3 \int_0^\infty dt_2 \nonumber\\
&\times &\int_0^\infty dt_1~ \frac{\cos(m_{1Q}t_1)\cos(m_{2Q}t_2)\cos(m_{3Q}t_3)\cos(m_{4Q}t_4)}{(t_1^2-x^2)^{\nu+1/2}(t_2^2-x^2)^{\mu+1/2}(t_3^2-x^2)^{\lambda+1/2}(t_4^2-x^2)^{\eta+1/2}}.\nonumber\\
\end{eqnarray}

We perform a Wick rotation   to go to the Euclidean space
and use again the Schwinger integral representation, we get
\begin{eqnarray}\label{e6}
T_4&=&\frac{(-i) 2^{\mu+\nu+\lambda+\eta}}{(\pi)^2 m_{1Q}^\nu m_{2Q}^\mu m_{3Q}^\lambda m_{4Q}^\eta \Gamma(\frac{n-\nu-\mu-\lambda-\eta}{2})}\int_0^1 du\int d^4x~e^{-iP.x}~A(u)\int_0^\infty dt_4\int_0^\infty dt_3 \int_0^\infty dt_2 \nonumber\\
&\times& \int_0^\infty dt_1\int_0^\infty dy_4\int_0^\infty dy_3 \int_0^\infty dy_2 \int_0^\infty dy_1 \int_0^\infty dy_0~
 y_0^{\frac{n-\nu-\mu-\lambda}{2}-1} e^{-y_0x^2}\nonumber\\
&\times&y_1^{\nu-\frac{1}{2}} e^{-y_1 (x^2+t_1^2)}y_2^{\mu-\frac{1}{2}} e^{-y_2 (x^2+t_2^2)}y_3^{\lambda-\frac{1}{2}} e^{-y_3 (x^2+t_3^2)} y_4^{\eta-\frac{1}{2}} e^{-y_4 (x^2+t_4^2)}\nonumber\\
&\times&\cos(m_{1Q}t_1)\cos(m_{2Q}t_2)\cos(m_{3Q}t_3)\cos(m_{4Q}t_4).\nonumber\\
\end{eqnarray}

The next step is to perform the Gaussian integral over four-$x$
and integrals over parameters $t_i$. As a result, we get
we get
\begin{eqnarray}\label{e10}
T_4&=&(-i) \frac{2^{\mu+\nu+\lambda+\eta-4}\pi^2}{ m_{1Q}^\nu m_{2Q}^\mu m_{3Q}^\lambda m_{4Q}^\eta \Gamma(\frac{n-\nu-\mu-\lambda-\eta}{2})}\int_0^1 du~A(u) \int_0^\infty dy_0~y_0^{\frac{n-\nu-\mu-\lambda-\eta}{2}-1}e^{\frac{-P^2}{4(y_0+y_1+y_2+y_3+y_4)}}\nonumber\\
&\times&\int_0^\infty dy_4 \int_0^\infty dy_3 \int_0^\infty dy_2 \int_0^\infty dy_1~\frac{y_1^{\nu-1} e^{-\frac{m^2_{1Q}}{4y_1}}y_2^{\mu-1} e^{-\frac{m^2_{2Q}}{4y_2}} y_3^{\lambda-1} e^{-\frac{m^2_{3Q}}{4y_3}}y_4^{\eta-1} e^{-\frac{m^2_{4Q}}{4y_4}}}{(y_0+y_1+y_2+y_3+y_4)^2}.
\end{eqnarray}

We  introduce the new set of variable as

\begin{eqnarray}\label{e11}
~~~~~~~~~~\rho =y_0+y_1+y_2+y_3+y_4,~~~~~~~~~~ v=\frac{y_1}{y_0+y_1+y_2+y_3+y_4},\nonumber\\
~~~~~~~~~~r=\frac{y_2}{y_0+y_1+y_2+y_3+y_4},~~~~~~~~~~w=\frac{y_3}{y_0+y_1+y_2+y_3+y_4},\nonumber\\
~~~~~~~~~~l=\frac{y_4}{y_0+y_1+y_2+y_3+y_4},~~~~~~~~~~~~~~~~~~~~~~~~~~~~~~~~~~~~~~~~~~~~~~
\end{eqnarray}
which leads to
\begin{eqnarray}\label{b12}
T_4&=&(-i) \frac{2^{\mu+\nu+\lambda+\eta-4}\pi^2}{ m_{1Q}^\nu m_{2Q}^\mu m_{3Q}^\lambda m_{4Q}^\eta \Gamma(\frac{n-\nu-\mu-\lambda-\eta}{2})}\int_0^1 du~A(u) \nonumber\\
&\times& \int_0^1 dw~ (\rho (1-v-r-l-w))^{\frac{n-\nu-\mu-\lambda-\eta}{2}-1}\int_0^\infty~\rho^2d\rho \int_0^1 dv\int_0^1 dr \nonumber\\
&\times& \int_0^1 dl (\rho v)^{\nu-1} e^{-\frac{m^2_{1Q}}{4\rho v}}(\rho r)^{\mu-1} e^{-\frac{m^2_{2Q}}{4\rho r}}(\rho w)^{\lambda-1} e^{-\frac{m^2_{3Q}}{4\rho w}}(\rho l)^{\eta-1} e^{-\frac{m^2_{4Q}}{4\rho l}}e^{\frac{-P^2}{4\rho}}.\nonumber\\
\end{eqnarray}

Applying the double Borel transformation and performing integral over the 
parameters $ u $ and $ \rho $, we obtain
\begin{eqnarray}\label{b14}
{\cal B}_{M_1}{\cal B}_{M_2}T_4&=&(-i) \frac{2^{-n}\pi^2}{ m_{1Q}^\nu m_{2Q}^\mu m_{3Q}^\lambda  m_{4Q}^\eta \Gamma(\frac{n-\nu-\mu-\lambda-\eta}{2}) }A(u_0)(M^2)^{\frac{n+\mu+\nu+\lambda+\eta}{2}}e^{\frac{q^2}{M_1^2+M_2^2}}\nonumber\\
&\times&\int_0^1 dv\int_0^1 dr \int_0^1 dl\int_0^1 dw~ (1-v-r-l-w)^{\frac{n-\nu-\mu-\lambda-\eta}{2}-1}\nonumber\\
 &\times& v^{\nu-1} e^{-\frac{m^2_{1Q}}{M^2 v}} r^{\mu-1} e^{-\frac{m^2_{2Q}}{M^2 r}} w^{\lambda-1} e^{-\frac{m^2_{3Q}}{M^2 w}} l^{\eta-1} e^{-\frac{m^2_{4Q}}{M^2 l}}.\nonumber\\
\end{eqnarray}

For further simplifications, we
 introduce the variables $ x $, $ y $, $ z $ and $t $, defined by

\begin{equation}\label{e15}
~~~~~~~~~~~v=zxt(1-y),~~~~~~~~r=zxyt,~~~~~~~~w=zt(1-x),~~~~~~~~l=z(1-t).
\end{equation}
Hence,
\begin{eqnarray}\label{e17}
{\cal B}_{M_1}{\cal B}_{M_2}T_4&=&(-i)  \frac{2^{-n}\pi^2}{ m_{1Q}^\nu m_{2Q}^\mu m_{3Q}^\lambda  m_{4Q}^\eta \Gamma(\frac{n-\nu-\mu-\lambda-\eta}{2}) }A(u_0)(M^2)^{\frac{n+\mu+\nu+\lambda+\eta}{2}}e^{\frac{q^2}{M_1^2+M_2^2}}\nonumber\\
&\times&\int_0^4 dz\int_0^1 dx \int_0^1 dy\int_0^1 dt~  z^{\nu+\mu+\lambda+\eta-1}(1-z)^{\frac{n-\nu-\mu-\lambda-\eta}{2}-1} t^{\nu+\mu+\lambda-1} (1-t)^{\eta-1}\nonumber\\
 &\times& x^{\nu+\mu-1} (1-x)^{\lambda-1}y^{\mu-1}(1-y)^{\nu-1} e^{-\frac{m^2_{1Q}}{M^2 zxt(1-y)}} e^{-\frac{m^2_{2Q}}{M^2 zxyt}}  e^{-\frac{m^2_{3Q}}{M^2 zt(1-x)}} e^{-\frac{m^2_{4Q}}{M^2 z(1-t)}}.\nonumber\\
\end{eqnarray}

\subsection*{\textit{Spectral representation and continuum subtraction  for  systems containing four heavy quarks}}
We again start with the  fallowing  general form:
\baeeq \label{e18}
\Pi_4&=&{\cal C}(M^2)^{\frac{n+\mu+\nu+\lambda+\eta}{2}} A(u_0)\int_0^4 dz\int_0^1 dx \int_0^1 dy\int_0^1 dt~  z^{\nu+\mu+\lambda+\eta-1}(1-z)^{\frac{n-\nu-\mu-\lambda-\eta}{2}-1} \nonumber\\
 &\times& t^{\nu+\mu+\lambda-1}(1-t)^{\eta-1} x^{\nu+\mu-1} (1-x)^{\lambda-1}y^{\mu-1}(1-y)^{\nu-1} e^{-\frac{m^2_{1Q}}{M^2 zxt(1-y)}}\nonumber\\
 &\times& e^{-\frac{m^2_{2Q}}{M^2 zxyt}}  e^{-\frac{m^2_{3Q}}{M^2 zt(1-x)}} e^{-\frac{m^2_{4Q}}{M^2 z(1-t)}}.
\eaeeq
where 
\baeeq
{\cal C}=(-i)  \frac{2^{-n}\pi^2}{ m_{1Q}^\nu m_{2Q}^\mu m_{3Q}^\lambda  m_{4Q}^\eta \Gamma(\frac{n-\nu-\mu-\lambda-\eta}{2}) }.
\eaeeq

As usual, the first step is to expand $A(u_0)$ in series,
which leads to
\baeeq \label{e20}
\Pi_4&=&{\cal C}\Bigg(\frac{M_1^2 M_2^2}{M_1^2+M_2^2}\Bigg)^{\frac{n+\mu+\nu+\lambda+\eta}{2}}\sum a_k\Bigg(\frac{ M_2^2}{M_1^2+M_2^2}\Bigg)^k\int_0^4 dz\int_0^1 dx \int_0^1 dy\int_0^1 dt~  z^{\nu+\mu+\lambda+\eta-1}\nonumber\\
 &\times&(1-z)^{\frac{n-\nu-\mu-\lambda-\eta}{2}-1} t^{\nu+\mu+\lambda-1} (1-t)^{\eta-1} x^{\nu+\mu-1} (1-x)^{\lambda-1}\nonumber\\
 &\times&y^{\mu-1}(1-y)^{\nu-1} e^{-\frac{m^2_{1Q}}{M^2 zxt(1-y)}} e^{-\frac{m^2_{2Q}}{M^2 zxyt}}  e^{-\frac{m^2_{3Q}}{M^2 zt(1-x)}} e^{-\frac{m^2_{4Q}}{M^2 z(1-t)}}.
\eaeeq

Now, the new variables, $\sigma_i=\frac{1}{M_i^2}$, are introduced. 
After some manipulations, we get

\baeeq \label{e21}
\Pi_4
&=&{\cal C}\sum_k a_k \frac{(-1)^k}{\Gamma(\frac{n+\mu+\nu+\lambda+\eta}{2}+k)}\int_0^4 dz\int_0^1 dx \int_0^1 dy\int_0^1 dt~  z^{\nu+\mu+\lambda+\eta-1}    (1-z)^{\frac{n-\nu-\mu-\lambda-\eta}{2}-1}\nonumber\\
 &\times&y^{\mu-1}(1-y)^{\nu-1} t^{\nu+\mu+\lambda-1} (1-t)^{\eta-1} x^{\nu+\mu-1} (1-x)^{\lambda-1}\nonumber\\
 &\times&\int_0^\infty dl~l^{\frac{n+\mu+\nu+\lambda+\eta}{2}+k-1}\Bigg((\frac{d}{dl})^ke^{-(l+\frac{m^2_{1Q}}{zxt(1-y)}+\frac{m^2_{2Q}}{ zxyt}+\frac{m^2_{3Q}}{ zt(1-x)}+\frac{m^2_{4Q}}{ z(1-t)})\sigma_1}\Bigg)\nonumber\\
 &\times&e^{-(l+\frac{m^2_{1Q}}{zxt(1-y)}+\frac{m^2_{2Q}}{ zxyt}+\frac{m^2_{3Q}}{ zt(1-x)}+\frac{m^2_{4Q}}{ z(1-t)})\sigma_2}.
\eaeeq

The double Borel transformation 
with respect to $\sigma_1\rar\frac{1}{s_1}$ and $\sigma_2\rar\frac{1}{s_2}$
are applied. From similar manner to the previous cases,
the spectral density is found as

\baeeq \label{e23}
\rho_4(s_1,s_2)&=&{\cal C}\sum_k a_k \frac{(-1)^k}{\Gamma(\frac{n+\mu+\nu+\lambda+\eta}{2}+k)}\int_0^4 dz\int_0^1 dx \int_0^1 dy\int_0^1 dt~  z^{\nu+\mu+\lambda+\eta-1}    (1-z)^{\frac{n-\nu-\mu-\lambda-\eta}{2}-1}\nonumber\\
 &\times&y^{\mu-1}(1-y)^{\nu-1} t^{\nu+\mu+\lambda-1} (1-t)^{\eta-1} x^{\nu+\mu-1} (1-x)^{\lambda-1}\nonumber\\
&\times &\int_0^\infty dl~l^{\frac{n+\mu+\nu+\lambda+\eta}{2}+k-1}\Bigg((\frac{d}{dl})^k\delta(s_1-(l+\frac{m^2_{1Q}}{zxt(1-y)}+\frac{m^2_{2Q}}{ zxyt}+\frac{m^2_{3Q}}{ zt(1-x)}+\frac{m^2_{4Q}}{ z(1-t)})\Bigg)\nonumber\\
&\times &\delta\Big(s_2-(l+\frac{m^2_{1Q}}{zxt(1-y)}+\frac{m^2_{2Q}}{ zxyt}+\frac{m^2_{3Q}}{ zt(1-x)}+\frac{m^2_{4Q}}{ z(1-t)})\Big).
\eaeeq

Performing the integration over $l$, finally we obtain the following expression for the double spectral density:
\baeeq \label{e24}
\rho_4(s_1,s_2)&=&{\cal C}\sum_k a_k \frac{(-1)^k}{\Gamma(\frac{n+\mu+\nu+\lambda+\eta}{2}+k)}\int_0^4 dz\int_0^1 dx \int_0^1 dy\int_0^1 dt~  z^{\nu+\mu+\lambda+\eta-1} \nonumber\\
 &\times&   (1-z)^{\frac{n-\nu-\mu-\lambda-\eta}{2}-1}y^{\mu-1}(1-y)^{\nu-1} t^{\nu+\mu+\lambda-1} (1-t)^{\eta-1} x^{\nu+\mu-1} (1-x)^{\lambda-1}\nonumber\\
&\times & \Bigg(s_1-\frac{m^2_{1Q}}{zxt(1-y)}-\frac{m^2_{2Q}}{ zxyt}-\frac{m^2_{3Q}}{ zt(1-x)}-\frac{m^2_{4Q}}{ z(1-t)}\Bigg)^{\frac{n+\mu+\nu+\lambda+\eta}{2}+k-1}\nonumber\\
&\times&\Bigg((\frac{d}{ds_1})^k\delta(s_2-s_1)\Bigg)\theta\Big(s_1-\frac{m^2_{1Q}}{zxt(1-y)}-\frac{m^2_{2Q}}{ zxyt}-\frac{m^2_{3Q}}{ zt(1-x)}-\frac{m^2_{4Q}}{ z(1-t)}\Big).
\eaeeq

Using this spectral density, the continuum subtracted correlation function in the Borel scheme corresponding to the considered term can be written as
\baeeq \label{e25} 
\Pi_4^{sub}(M_1^2,M_2^2)=\int_{s_L}^{s_0}ds_1\int_{s_L}^{s_0}ds_2~\rho(s_1,s_2)e^{-s_1/M_1^2}e^{-s_2/M_2^2},
\eaeeq
where $ s_L=(m_{1Q}+m_{2Q}+m_{3Q}+m_{4Q})^2 $.
Defining new variables, $s_1=2 s v$ and $s_2=2 s(1- v)$, we get
\eAPP
\baeeq \label{e26}
\Pi_4^{sub}(M_1^2,M_2^2)=\int_{s_L}^{s_0} ds\int_0^1 dv~\rho(s_1,s_2)(4s)e^{-2sv/M_1^2}e^{-2s(1-v)/M_2^2}.
\eaeeq

Using the expression for the spectral density, one can get
\baeeq \label{e27}
\Pi_4^{sub}(M_1^2,M_2^2)&=&{\cal C}\sum_k a_k \frac{(-1)^k}{\Gamma(\frac{n+\mu+\nu+\lambda+\eta}{2}+k)}\int_{s_L}^{s_0} ds\int_0^1 dv\frac{1}{2^ks^k}\Bigg((\frac{d}{dv})^k\delta(v-1/2)\Bigg)\nonumber\\
&\times&\int_0^4 dz\int_0^1 dx \int_0^1 dy\int_0^1 dt~  z^{\nu+\mu+\lambda+\eta-1}    (1-z)^{\frac{n-\nu-\mu-\lambda-\eta}{2}-1}\nonumber\\
 &\times&y^{\mu-1}(1-y)^{\nu-1} t^{\nu+\mu+\lambda-1} (1-t)^{\eta-1} x^{\nu+\mu-1} (1-x)^{\lambda-1}\nonumber\\
 &\times&\Bigg(2sv-\frac{m^2_{1Q}}{zxt(1-y)}-\frac{m^2_{2Q}}{ zxyt}-\frac{m^2_{3Q}}{ zt(1-x)}-\frac{m^2_{4Q}}{ z(1-t)}\Bigg)^{\frac{n+\mu+\nu+\lambda+\eta}{2}+k-1}\nonumber\\
&\times&\theta\Big(2sv-\frac{m^2_{1Q}}{zxt(1-y)}-\frac{m^2_{2Q}}{ zxyt}-\frac{m^2_{3Q}}{ zt(1-x)}-\frac{m^2_{4Q}}{ z(1-t)}\Big)e^{-2sv/M_1^2}e^{-2s(1-v)/M_2^2}.\nonumber\\
\eaeeq \label{e28}
Integrating over $v$, finally leads to
\baeeq
\Pi_4^{sub}(M_1^2,M_2^2)&=&(-i) \sum_k a_k \frac{2^{-n}\pi^2}{ m_{1Q}^\nu m_{2Q}^\mu m_{3Q}^\lambda  m_{4Q}^\eta \Gamma(\frac{n-\nu-\mu-\lambda-\eta}{2})\Gamma(\frac{n+\mu+\nu+\lambda+\eta}{2}+k)}\int ds\frac{1}{2^ks^k} \nonumber\\
&\times &\Bigg[(\frac{d}{dv})^k    
\int_0^4 dz~\theta(2sv-\frac{m^2_{1Q}}{zxt(1-y)}-\frac{m^2_{2Q}}{ zxyt}-\frac{m^2_{3Q}}{ zt(1-x)}-\frac{m^2_{4Q}}{ z(1-t)})\nonumber\\
&\times&\int_0^1 dx \int_0^1 dy\int_0^1 dt~  z^{\nu+\mu+\lambda+\eta-1}    (1-z)^{\frac{n-\nu-\mu-\lambda-\eta}{2}-1}\nonumber\\
 &\times&y^{\mu-1}(1-y)^{\nu-1} t^{\nu+\mu+\lambda-1} (1-t)^{\eta-1} x^{\nu+\mu-1} (1-x)^{\lambda-1}\nonumber\\
 &\times&\Bigg(2sv-\frac{m^2_{1Q}}{zxt(1-y)}-\frac{m^2_{2Q}}{ zxyt}-\frac{m^2_{3Q}}{ zt(1-x)}-\frac{m^2_{4Q}}{ z(1-t)}\Bigg)^{\frac{n+\mu+\nu+\lambda+\eta}{2}+k-1}\nonumber\\
 &\times &e^{-2sv/M_1^2}e^{-2s(1-v)/M_2^2}\Bigg]_{v=1/2}.
\eaeeq

\subsection{Systems containing  five heavy quarks}
The required function to be evaluated in this case is
\begin{eqnarray}\label{f1}
T_5&=& \int_0^1 du\int d^4x e^{iP.x}\frac{ A(u)}{(\sqrt{-x^2})^n}\nonumber\\
&\times&K_\nu(m_{1Q}\sqrt{-x^2})K_\mu(m_{2Q}\sqrt{-x^2})K_\lambda(m_{3Q}\sqrt{-x^2})K_\eta(m_{4Q}\sqrt{-x^2})K_\xi(m_{5Q}\sqrt{-x^2}),\nonumber\\
\end{eqnarray}
where $ m_{iQ} $, are the masses of heavy quarks. Using the  integral representation
 of the modified Bessel function,
we have
\begin{eqnarray}\label{f3}
T_5&=&\frac{\Gamma(\nu+ 1/2)\Gamma(\mu+ 1/2)\Gamma(\lambda+ 1/2)\Gamma(\eta+ 1/2)\Gamma(\xi+ 1/2) ~2^{\mu+\nu+\lambda+\eta+\xi}}{(\pi)^{5/2} m_{1Q}^\nu m_{2Q}^\mu m_{3Q}^\lambda m_{4Q}^\eta m_{5Q}^\xi}\nonumber\\
&\times&\int_0^\infty dt_5\int_0^\infty dt_4\int_0^\infty dt_3 \int_0^\infty dt_2 \int_0^\infty dt_1 \int_0^1 du\int d^4x~e^{iP.x}~A(u) \nonumber\\
&\times& \frac{\cos(m_{1Q}t_1)\cos(m_{2Q}t_2)\cos(m_{3Q}t_3)\cos(m_{4Q}t_4)\cos(m_{5Q}t_5)}{(t_1^2-x^2)^{\nu+1/2}(t_2^2-x^2)^{\mu+1/2}(t_3^2-x^2)^{\lambda+1/2}(t_4^2-x^2)^{\eta+1/2}
(t_5^2-x^2)^{\xi+1/2}(-x^2)^{\frac{n-\mu-\nu-\lambda-\eta-\xi}{2}}}.\nonumber\\
\end{eqnarray}

In this step, for further calculations, we again perform a Wick rotation   to go the Euclidean space
and use the  Schwinger representation. We get
\begin{eqnarray}\label{f6}
T_5&=&\frac{(-i) 2^{\mu+\nu+\lambda+\eta+\xi}}{(\pi)^{5/2} m_{1Q}^\nu m_{2Q}^\mu m_{3Q}^\lambda m_{4Q}^\eta m_{5Q}^\xi \Gamma(\frac{n-\mu-\nu-\lambda-\eta-\xi}{2})}\int_0^1 du\int d^4xe^{-iP.x}~A(u)\int_0^\infty dt_5\int_0^\infty dt_4\int_0^\infty dt_3 \nonumber\\
&\times& \int_0^\infty dt_2 \int_0^\infty dt_1\int_0^\infty dy_5\int_0^\infty dy_4\int_0^\infty dy_3 \int_0^\infty dy_2 \int_0^\infty dy_1 \int_0^\infty dy_0~y_0^{\frac{n-\mu-\nu-\lambda-\eta-\xi}{2}-1} e^{-y_0 x^2}\nonumber\\
&\times&y_1^{\nu-\frac{1}{2}} e^{-y_1 (x^2+t_1^2)}y_2^{\mu-\frac{1}{2}} e^{-y_2 (x^2+t_2^2)}y_3^{\lambda-\frac{1}{2}} e^{-y_3 (x^2+t_3^2)}y_4^{\eta-\frac{1}{2}} e^{-y_4 (x^2+t_4^2)}y_5^{\xi-\frac{1}{2}} e^{-y_5 (x^2+t_5^2)}  \nonumber\\
&\times&\cos(m_{1Q}t_1)\cos(m_{2Q}t_2)\cos(m_{3Q}t_3)\cos(m_{4Q}t_4)\cos(m_{5Q}t_5).\nonumber\\
\end{eqnarray}

The next step is to perform the Gaussian integral over $ x $
and integrate over $ t_i $'s. This leads to
\begin{eqnarray}\label{f10}
T_5&=&(-i) \frac{2^{\mu+\nu+\lambda+\eta+\xi-5}\pi^2}{ m_{1Q}^\nu m_{2Q}^\mu m_{3Q}^\lambda m_{4Q}^\eta m_{5Q}^\xi \Gamma(\frac{n-\mu-\nu-\lambda-\eta-\xi}{2})}\int_0^1 du~A(u) \int_0^\infty dy_5\int_0^\infty dy_4 \int_0^\infty dy_3  \nonumber\\
&\times&\int_0^\infty dy_2\int_0^\infty dy_1\int_0^\infty dy_0~e^{\frac{-P^2}{4(y_0+y_1+y_2+y_3+y_4+y_5)}} y_0^{\frac{n-\mu-\nu-\lambda-\eta-\xi}{2}-1}\nonumber\\
&\times&\frac{y_1^{\nu-1} e^{-\frac{m^2_{1Q}}{4y_1}}y_2^{\mu-1} e^{-\frac{m^2_{2Q}}{4y_2}} y_3^{\lambda-1} e^{-\frac{m^2_{3Q}}{4y_3}}y_4^{\eta-1} e^{-\frac{m^2_{4Q}}{4y_4}}y_5^{\xi-1} e^{-\frac{m^2_{5Q}}{4y_5}}}{(y_0+y_1+y_2+y_3+y_4+y_5)^2}. \nonumber\\
\end{eqnarray}

We  introduce the variables $ \rho $, $ v $, $ l $, $ r $, $ h $ and $ w $, defined by

\begin{eqnarray}\label{f11}
~~~~~~~~~~\rho&=&y_1+y_2+y_3+y_4+y_5,~~~~~~~~~~v=\frac{y_1}{y_1+y_2+y_3+y_4+y_5},\nonumber\\
r&=&\frac{y_2}{y_1+y_2+y_3+y_4},~~~~~~~~~~~~~~~~w=\frac{y_3}{y_1+y_2+y_3+y_4+y_5},\nonumber\\
l&=&\frac{y_4}{y_1+y_2+y_3+y_4+y_5},~~~~~~~~~~h=\frac{y_5}{y_1+y_2+y_3+y_4+y_5},
\end{eqnarray}
to write the function $ T_5 $ as
\begin{eqnarray}\label{f12}
T_5&=&(-i) \frac{2^{\mu+\nu+\lambda+\eta+\xi-5}\pi^2}{ m_{1Q}^\nu m_{2Q}^\mu m_{3Q}^\lambda m_{4Q}^\eta m_{5Q}^\xi \Gamma(\frac{n-\mu-\nu-\lambda-\eta-\xi}{2}) }\int_0^1 du~A(u) \int_0^1 dw \int_0^1 dv \int_0^1  dr 
\nonumber\\
&\times &\int_0^1 dl\int_0^1 dh\int_0^\infty\rho^3d\rho~e^{\frac{-P^2}{4\rho}}(\rho(1-v-r-l-w-h))^{\frac{n-\mu-\nu-\lambda-\eta-\xi}{2}-1}\nonumber\\
&\times &(\rho v)^{\nu-1} e^{-\frac{m^2_{1Q}}{4\rho v}}(\rho r)^{\mu-1} e^{-\frac{m^2_{2Q}}{4\rho r}}(\rho w)^{\lambda-1} e^{-\frac{m^2_{3Q}}{4\rho w}}(\rho l)^{\eta-1} e^{-\frac{m^2_{4Q}}{4\rho l}}(\rho h)^{\xi-1} e^{-\frac{m^2_{5Q}}{4\rho h}}.\nonumber\\
\end{eqnarray}

Now we apply the double Borel transformation
and perform the integrals over $ u $ and $ \rho $,
\begin{eqnarray}\label{f14}
{\cal B}_{M_1}{\cal B}_{M_2}T_5&=&(-i) \frac{2^{-1-n}\pi^2}{ m_{1Q}^\nu m_{2Q}^\mu m_{3Q}^\lambda m_{4Q}^\eta m_{5Q}^\xi \Gamma(\frac{n-\mu-\nu-\lambda-\eta-\xi}{2})}A(u_0)(M^2)^{\frac{n+\mu+\nu+\lambda+\eta+\xi}{2}}\nonumber\\
 &\times&\int_0^1 dw \int_0^1 dv \int_0^1  dr \int_0^1 dl \int_0^1 dh(1-v-r-l-w-h)^{\frac{n-\mu-\nu-\lambda-\eta-\xi}{2}-1}  \nonumber\\
&\times&v^{\nu-1}e^{-\frac{m^2_{1Q}}{M^2v}} r^{\mu-1} e^{-\frac{m^2_{2Q}}{M^2 r}}  w^{\lambda-1} e^{-\frac{m^2_{3Q}}{M^2 w}}l^{\eta-1} e^{-\frac{m^2_{4Q}}{M^2 l}} h^{\xi-1} e^{-\frac{m^2_{5Q}}{M^2 h}}e^{\frac{q^2}{M_1^2+M_2^2}}.\nonumber\\
\end{eqnarray}

The following new set of variables make the
function $ {\cal B}_{M_1}{\cal B}_{M_2}T $ easy to process:
\begin{eqnarray}\label{f15}
~~~v=zxtj(1-y),~~r=zxytj,~~w=ztj(1-x),~~l=zj(1-t),~~h=z(1-j),
\end{eqnarray}
so,we get
\begin{eqnarray}\label{f16}
{\cal B}_{M_1}{\cal B}_{M_2}T_5&=&(-i) \frac{2^{-1-n}\pi^2}{ m_{1Q}^\nu m_{2Q}^\mu m_{3Q}^\lambda m_{4Q}^\eta m_{5Q}^\xi \Gamma(\frac{n-\mu-\nu-\lambda-\eta-\xi}{2})}A(u_0)(M^2)^{\frac{n+\mu+\nu+\lambda+\eta+\xi}{2}}\nonumber\\
 &\times&\int_0^5 dz\int_0^1 dx \int_0^1 dy\int_0^1 dt \int_0^1 dl \int_0^1 dj~ z^{\nu+\mu+\lambda+\eta+\xi-1}(1-z)^{\frac{n-\nu-\mu-\lambda-\eta-\xi}{2}-1}\nonumber\\
 &\times& j^{\nu+\mu+\lambda+\eta-1} (1-j)^{\xi-1} t^{\nu+\mu+\lambda-1} (1-t)^{\eta-1}  x^{\nu+\mu-1} (1-x)^{\lambda-1}y^{\mu-1}(1-y)^{\nu-1}\nonumber\\
 &\times&e^{-\frac{m^2_{1Q}}{M^2zxtj(1-y)}} e^{-\frac{m^2_{2Q}}{M^2 zxytj}} e^{-\frac{m^2_{3Q}}{M^2 ztj(1-x)}} e^{-\frac{m^2_{4Q}}{M^2 zj(1-t)}}e^{-\frac{m^2_{5Q}}{M^2 z(1-j)}}.
\end{eqnarray}

\subsection*{\textit{Spectral representation and continuum subtraction for  systems containing five heavy quarks}}

We take the general  term  
\baeeq \label{f17}
\Pi_5&=&{\cal C}(M^2)^{\frac{n+\mu+\nu+\lambda+\eta+\xi}{2}} A(u_0)\int_0^5 dz\int_0^1 dx \int_0^1 dy\int_0^1 dt\nonumber\\
 &\times&\int_0^1 dj~ z^{\nu+\mu+\lambda+\eta+\xi-1}(1-z)^{\frac{n-\nu-\mu-\lambda-\eta-\xi}{2}-1}\nonumber\\
 &\times& j^{\nu+\mu+\lambda+\eta-1} (1-j)^{\xi-1} t^{\nu+\mu+\lambda-1} (1-t)^{\eta-1}  x^{\nu+\mu-1} (1-x)^{\lambda-1}y^{\mu-1}(1-y)^{\nu-1}\nonumber\\
 &\times&e^{-\frac{m^2_{1Q}}{M^2zxtj(1-y)}} e^{-\frac{m^2_{2Q}}{M^2 zxytj}} e^{-\frac{m^2_{3Q}}{M^2 ztj(1-x)}} e^{-\frac{m^2_{4Q}}{M^2 zj(1-t)}}e^{-\frac{m^2_{5Q}}{M^2 z(1-j)}}, \nonumber\\
\eaeeq
where
\baeeq \label{f18}
{\cal C}=(-i) \frac{2^{-1-n}\pi^2}{ m_{1Q}^\nu m_{2Q}^\mu m_{3Q}^\lambda m_{4Q}^\eta m_{5Q}^\xi \Gamma(\frac{n-\mu-\nu-\lambda-\eta-\xi}{2})}.
\eaeeq
With expanding  $ A(u_0) $, this leads to

\baeeq \label{f20}
\Pi_5
&=&{\cal C}\sum_k a_k \frac{(-1)^k}{\Gamma(\frac{n+\mu+\nu+\lambda+\eta+\xi}{2}+k)}\int_0^5 dz\int_0^1 dx \int_0^1 dy\int_0^1 dt\nonumber\\
 &\times&\int_0^1 dj~ z^{\nu+\mu+\lambda+\eta+\xi-1}(1-z)^{\frac{n-\nu-\mu-\lambda-\eta-\xi}{2}-1} j^{\nu+\mu+\lambda+\eta-1} (1-j)^{\xi-1}\nonumber\\
 &\times& t^{\nu+\mu+\lambda-1} (1-t)^{\eta-1}  x^{\nu+\mu-1} (1-x)^{\lambda-1}y^{\mu-1}(1-y)^{\nu-1}\nonumber\\
 &\times&\int_0^\infty dl~l^{\frac{n+\mu+\nu+\lambda+\eta+\xi}{2}+k-1} e^{-(l+\frac{m^2_{1Q}}{zxtj(1-y)}+\frac{m^2_{2Q}}{ zxytj}+\frac{m^2_{3Q}}{ztj(1-x)}+\frac{m^2_{4Q}}{zj(1-t)}+\frac{m^2_{5Q}}{ z(1-j)})\sigma_2}\nonumber\\
 &\times& \Bigg((\frac{d}{dl})^ke^{-(l+\frac{m^2_{1Q}}{zxtj(1-y)}+\frac{m^2_{2Q}}{ zxytj}+\frac{m^2_{3Q}}{ztj(1-x)}+\frac{m^2_{4Q}}{zj(1-t)}+\frac{m^2_{5Q}}{ z(1-j)})\sigma_1}\Bigg).\nonumber\\
\eaeeq
By applying double Borel transformation with respect to $\sigma_1\rar\frac{1}{\tau_1}$ and $\sigma_2\rar\frac{1}{\tau_2}$, we obtain
\baeeq \label{f21}
{\cal B}_{1/\tau_1}{\cal B}_{1/\tau_2}\Pi_5&=&{\cal C}\sum_k a_k \frac{(-1)^k}{\Gamma(\frac{n+\mu+\nu+\lambda+\eta+\xi}{2}+k)}\int_0^\infty dl~l^{\frac{n+\mu+\nu+\lambda+\eta+\xi}{2}+k-1}
\nonumber\\
&\times&\int_0^5 dz\int_0^1 dx \int_0^1 dy\int_0^1 dt\int_0^1 dj~ z^{\nu+\mu+\lambda+\eta+\xi-1}(1-z)^{\frac{n-\nu-\mu-\lambda-\eta-\xi}{2}-1}\nonumber\\
 &\times& j^{\nu+\mu+\lambda+\eta-1} (1-j)^{\xi-1} t^{\nu+\mu+\lambda-1} (1-t)^{\eta-1}  x^{\nu+\mu-1} (1-x)^{\lambda-1}y^{\mu-1}(1-y)^{\nu-1}\nonumber\\
&\times&\Bigg((\frac{d}{dl})^k\delta(\tau_1-(l+\frac{m^2_{1Q}}{zxtj(1-y)}+\frac{m^2_{2Q}}{ zxytj}+\frac{m^2_{3Q}}{ztj(1-x)}+\frac{m^2_{4Q}}{zj(1-t)}+\frac{m^2_{5Q}}{ z(1-j)})\Bigg)\nonumber\\
&\times&\delta\Big(\tau_2-(l+\frac{m^2_{1Q}}{zxtj(1-y)}+\frac{m^2_{2Q}}{ zxytj}+\frac{m^2_{3Q}}{ztj(1-x)}+\frac{m^2_{4Q}}{zj(1-t)}+\frac{m^2_{5Q}}{ z(1-j)})\Big),\nonumber\\
\eaeeq
and this leads to the spectral density
\baeeq \label{f22}
\rho_5(s_1,s_2)&=&{\cal C}\sum_k a_k \frac{(-1)^k}{\Gamma(\frac{n+\mu+\nu+\lambda+\eta+\xi}{2}+k)}\int_0^\infty dl~l^{\frac{n+\mu+\nu+\lambda+\eta+\xi}{2}+k-1}
\nonumber\\
&\times&\int_0^5 dz\int_0^1 dx \int_0^1 dy\int_0^1 dt\int_0^1 dj~ z^{\nu+\mu+\lambda+\eta+\xi-1}(1-z)^{\frac{n-\nu-\mu-\lambda-\eta-\xi}{2}-1}\nonumber\\
 &\times& j^{\nu+\mu+\lambda+\eta-1} (1-j)^{\xi-1} t^{\nu+\mu+\lambda-1} (1-t)^{\eta-1}  x^{\nu+\mu-1} (1-x)^{\lambda-1}y^{\mu-1}(1-y)^{\nu-1}\nonumber\\
&\times&\Bigg((\frac{d}{dl})^k\delta(s_1-(l+\frac{m^2_{1Q}}{zxtj(1-y)}+\frac{m^2_{2Q}}{ zxytj}+\frac{m^2_{3Q}}{ztj(1-x)}+\frac{m^2_{4Q}}{zj(1-t)}+\frac{m^2_{5Q}}{ z(1-j)})\Bigg)\nonumber\\
&\times&\delta\Big(s_2-(l+\frac{m^2_{1Q}}{zxtj(1-y)}+\frac{m^2_{2Q}}{ zxytj}+\frac{m^2_{3Q}}{ztj(1-x)}+\frac{m^2_{4Q}}{zj(1-t)}+\frac{m^2_{5Q}}{ z(1-j)})\Big).\nonumber\\
\eaeeq

Performing the integration over $l$, finally we obtain the following expression for the double spectral density:
\baeeq \label{f23}
\rho_5(s_1,s_2)&=&{\cal C}\sum_k a_k \frac{(-1)^k}{\Gamma(\frac{n+\mu+\nu+\lambda+\eta+\xi}{2}+k)}\int_0^5 dz\int_0^1 dx \int_0^1 dy\int_0^1 dt\nonumber\\
&\times&\int_0^1 dj~ z^{\nu+\mu+\lambda+\eta+\xi-1}(1-z)^{\frac{n-\nu-\mu-\lambda-\eta-\xi}{2}-1}\nonumber\\
 &\times& j^{\nu+\mu+\lambda+\eta-1} (1-j)^{\xi-1} t^{\nu+\mu+\lambda-1} (1-t)^{\eta-1}  x^{\nu+\mu-1} (1-x)^{\lambda-1}y^{\mu-1}(1-y)^{\nu-1}\nonumber\\
&\times& \Bigg(s_1-(\frac{m^2_{1Q}}{zxtj(1-y)}+\frac{m^2_{2Q}}{ zxytj}+\frac{m^2_{3Q}}{ztj(1-x)}+\frac{m^2_{4Q}}{zj(1-t)}+\frac{m^2_{5Q}}{ z(1-j)})\Bigg)^{\frac{n+\mu+\nu+\lambda+\eta+\xi}{2}+k-1}\nonumber\\
&\times&\theta\Big(s_1-(\frac{m^2_{1Q}}{zxtj(1-y)}+\frac{m^2_{2Q}}{ zxytj}+\frac{m^2_{3Q}}{ztj(1-x)}+\frac{m^2_{4Q}}{zj(1-t)}+\frac{m^2_{5Q}}{ z(1-j)})\Big)\nonumber\\
&\times&\Bigg((\frac{d}{ds_1})^k\delta(s_2-s_1)\Bigg).
\eaeeq

Using this spectral density, the continuum subtracted correlation function in the Borel scheme corresponding to the considered term can be written as:
\baeeq \label{f24}
\Pi_5^{sub}(M_1^2,M_2^2)=\int_{s_L}^{s_0}ds_1\int_{s_L}^{s_0}ds_2~\rho(s_1,s_2)e^{-s_1/M_1^2}e^{-s_2/M_2^2},
\eaeeq
where $ s_L=(m_{1Q}+m_{2Q}+m_{3Q}+m_{4Q}+m_{5Q})^2 $.
Defining new variables, $s_1=2 s v$ and $s_2=2 s(1- v)$, we get
\baeeq \label{f25}
\Pi_5^{sub}(M_1^2,M_2^2)=\int_{s_L}^{s_0} ds\int_0^1 dv~\rho(s_1,s_2)(4s)e^{-2sv/M_1^2}e^{-2s(1-v)/M_2^2}.
\eaeeq

Using the expression for the spectral density, one can get
\baeeq \label{f26}
\Pi_5^{sub}(M_1^2,M_2^2)&=&{\cal C}\sum_k a_k \frac{(-1)^k}{\Gamma(\frac{n+\mu+\nu+\lambda+\eta+\xi}{2}+k)}\int_{s_L}^{s_0} ds\nonumber\\
&\times& \int_0^1 dv\frac{1}{2^ks^k}\Bigg((\frac{d}{dv})^k\delta(v-1/2)\Bigg) e^{-2sv/M_1^2}e^{-2s(1-v)/M_2^2}\nonumber\\
&\times&\int_0^5 dz\int_0^1 dx \int_0^1 dy\int_0^1 dt\int_0^1 dj~ z^{\nu+\mu+\lambda+\eta+\xi-1}(1-z)^{\frac{n-\nu-\mu-\lambda-\eta-\xi}{2}-1}\nonumber\\
 &\times& j^{\nu+\mu+\lambda+\eta-1} (1-j)^{\xi-1} t^{\nu+\mu+\lambda-1} (1-t)^{\eta-1}  x^{\nu+\mu-1} (1-x)^{\lambda-1}y^{\mu-1}(1-y)^{\nu-1}\nonumber\\
 &\times&\Bigg(2sv-(\frac{m^2_{1Q}}{zxtj(1-y)}+\frac{m^2_{2Q}}{ zxytj}+\frac{m^2_{3Q}}{ztj(1-x)}+\frac{m^2_{4Q}}{zj(1-t)}+\frac{m^2_{5Q}}{ z(1-j)})\Bigg)^{\frac{n+\mu+\nu+\lambda+\eta+\xi}{2}+k-1}\nonumber\\
&\times&\theta\Big(2sv-(\frac{m^2_{1Q}}{zxtj(1-y)}+\frac{m^2_{2Q}}{ zxytj}+\frac{m^2_{3Q}}{ztj(1-x)}+\frac{m^2_{4Q}}{zj(1-t)}+\frac{m^2_{5Q}}{ z(1-j)})\Big).\nonumber\\
\eaeeq

By integrating over $v$, finally we obtain
\baeeq \label{f27}
\Pi_5^{sub}(M_1^2,M_2^2)&=&\sum_k a_k \frac{(-i)2^{-1-n}\pi^2}{m_{1Q}^\nu m_{2Q}^\mu m_{3Q}^\lambda m_{4Q}^\eta m_{5Q}^\xi \Gamma(\frac{n-\mu-\nu-\lambda-\eta-\xi}{2})\Gamma(\frac{n+\mu+\nu+\lambda+\eta+\xi}{2}+k)}\int_{s_L}^{s_0}  ds\frac{1}{2^ks^k} \nonumber\\
&\times&\Bigg[(\frac{d}{dv})^k    
\int_0^5 dz~\theta\Big(2sv-(\frac{m^2_{1Q}}{zxtj(1-y)}+\frac{m^2_{2Q}}{ zxytj}+\frac{m^2_{3Q}}{ztj(1-x)}+\frac{m^2_{4Q}}{zj(1-t)}+\frac{m^2_{5Q}}{ z(1-j)})\Big)\nonumber\\
&\times& \int_0^1 dx \int_0^1 dy\int_0^1 dt\int_0^1 dj~ z^{\nu+\mu+\lambda+\eta+\xi-1}(1-z)^{\frac{n-\nu-\mu-\lambda-\eta-\xi}{2}-1}\nonumber\\
&\times&  j^{\nu+\mu+\lambda+\eta-1} (1-j)^{\xi-1} t^{\nu+\mu+\lambda-1} (1-t)^{\eta-1}  x^{\nu+\mu-1} (1-x)^{\lambda-1}y^{\mu-1}(1-y)^{\nu-1}\nonumber\\
&\times& \Bigg(2sv-(\frac{m^2_{1Q}}{zxtj(1-y)}+\frac{m^2_{2Q}}{ zxytj}+\frac{m^2_{3Q}}{ztj(1-x)}+\frac{m^2_{4Q}}{zj(1-t)}+\frac{m^2_{5Q}}{ z(1-j)})\Bigg)^{\frac{n+\mu+\nu+\lambda+\eta+\xi}{2}+k-1}\nonumber\\
&\times& e^{-2sv/M_1^2}e^{-2s(1-v)/M_2^2}\Bigg]_{v=1/2}.
\eaeeq
\section{Conclusion}\label{concl} 
We worked out the mathematics required for the calculations
of the parameters related to the spectroscopy  as well as  the 
electromagnetic, weak and strong decays of the light and heavy 
systems with two-five heavy $ b $ or $ c $ quarks. In particular, we 
presented the calculations required in the Fourier and Borel 
transformations as well as continuum subtraction of the considered 
systems. We showed that by choosing an appropriate representation 
of the modified Bessel functions of the  second kind and applying 
successive Borel transformations
with the aim of more suppression of the unwanted contributions,
 how we can get 
finite results without any divergence. 
Such divergences are appeared for systems of  two heavy quarks by choosing a special integral form 
of the modified Bessel functions and performing the regular calculations 
according to the standard prescriptions of the QCD sum
rule approach as done 
 in the Appendix C of the Ref.~\cite{Ozdem:2017jqh}.

The method presented here greatly simplifies the 
calculations of the correlation functions for systems containing more than two heavy quarks.
Our results can be used in calculations of many parameters of
 the conventional and non-conventional heavy hadrons and their interactions with other particles using the 
 non-perturbative approaches like QCD sum rules. 

\section{Acknowledgments}
\vspace{-0.2cm} S. R. and A.~R. O. would like to thank School of Physics, Institute for
Research in Fundamental Sciences (IPM)
 for their hospitality where a part of this work was performed. K.~A. thanks Do\v{g}u\c{s} University for the partial financial support
through the grant BAP 2015-16-D1-B04.

\section{Appendix: DAs of pseudoscalar mesons}
In this Appendix, we present the matrix elements $\lla {\cal P}(q) \vel \bar{q}(x)
\Gamma q (0) \ver 0 \rra$ and $\lla {\cal P}(q) \vel \bar{q}(x) \Gamma G_{\mu\nu} q(0)
\ver 0 \rra$ representing the interactions of the particles under consideration in terms of the wave functions of pseudoscalar mesons \cite{R10511,R10512,R10513}:
\bea
\label{e10527}
\lla {\cal P}(q)\vel \bar q(x) \gamma_\mu \gamma_5 q(0)\ver 0 \rra \es
-i f_{\cal P} q_\mu  \int_0^1 du  e^{i \bar u q x}
    \left( \varphi_{\cal P}(u) + {1\over 16} m_{\cal P}^2
x^2 {\Bbb{A}}(u) \right) \nnb \\
\ek {i\over 2} f_{\cal P} m_{\cal P}^2 {x_\mu\over qx}
\int_0^1 du e^{i \bar u qx} {\Bbb{B}}(u)~,\nnb \\
\lla {\cal P}(q)\vel \bar q(x) i \gamma_5 q(0)\ver 0 \rra \es
\mu_{\cal P} \int_0^1 du e^{i \bar u qx} \varphi_P(u)~,\nnb \\
\lla {\cal P}(q)\vel \bar q(x) \sigma_{\alpha \beta} \gamma_5 q(0)\ver 0 \rra \es
{i\over 6} \mu_{\cal P} \left( 1 - \widetilde{\mu}_{\cal P}^2 \right)
\left( q_\alpha x_\beta - q_\beta x_\alpha\right)
\int_0^1 du e^{i \bar u qx} \varphi_\sigma(u)~,\nnb \\
\lla {\cal P}(q)\vel \bar q(x) \sigma_{\mu \nu} \gamma_5 g_s
G_{\alpha \beta}(v x) q(0)\ver 0 \rra \es i \mu_{\cal P} \left[
q_\alpha q_\mu \left( g_{\nu \beta} - {1\over qx}(q_\nu x_\beta +
q_\beta x_\nu) \right) \right. \nnb \\
\ek q_\alpha q_\nu \left( g_{\mu \beta} -
{1\over qx}(q_\mu x_\beta + q_\beta x_\mu) \right) \nnb \\
\ek q_\beta q_\mu \left( g_{\nu \alpha} - {1\over qx}
(q_\nu x_\alpha + q_\alpha x_\nu) \right) \nnb \\
\ar q_\beta q_\nu \left. \left( g_{\mu \alpha} -
{1\over qx}(q_\mu x_\alpha + q_\alpha x_\mu) \right) \right] \nnb \\
\cp \int  D \alpha e^{i (\alpha_{\bar q} +
v \alpha_g) qx} {\cal T}(\alpha_i)~,\nnb \\
\lla {\cal P}(q)\vel \bar q(x) \gamma_\mu \gamma_5 g_s
G_{\alpha \beta} (v x) q(0)\ver 0 \rra \es q_\mu (q_\alpha x_\beta -
q_\beta x_\alpha) {1\over qx} f_{\cal P} m_{\cal P}^2
\int  D\alpha e^{i (\alpha_{\bar q} + v \alpha_g) qx}
{\cal A}_\parallel (\alpha_i) \nnb \\
\ar \left[q_\beta \left( g_{\mu \alpha} - {1\over qx}
(q_\mu x_\alpha + q_\alpha x_\mu) \right) \right. \nnb \\
\ek q_\alpha \left. \left(g_{\mu \beta}  - {1\over qx}
(q_\mu x_\beta + q_\beta x_\mu) \right) \right]
f_{\cal P} m_{\cal P}^2 \nnb \\
\cp \int D\alpha e^{i (\alpha_{\bar q} + v \alpha _g)
q x} {\cal A}_\perp(\alpha_i)~,\nnb \\
\lla {\cal P}(q)\vel \bar q(x) \gamma_\mu i g_s G_{\alpha \beta}
(v x) q(0)\ver 0 \rra \es q_\mu (q_\alpha x_\beta - q_\beta x_\alpha)
{1\over qx} f_{\cal P} m_{\cal P}^2 \int  D\alpha e^{i (\alpha_{\bar q} +
v \alpha_g) qx} {\cal V}_\parallel (\alpha_i) \nnb \\
\ar \left[q_\beta \left( g_{\mu \alpha} - {1\over qx}
(q_\mu x_\alpha + q_\alpha x_\mu) \right) \right. \nnb \\
\ek q_\alpha \left. \left(g_{\mu \beta}  - {1\over qx}
(q_\mu x_\beta + q_\beta x_\mu) \right) \right] f_{\cal P} m_{\cal P}^2 \nnb \\
    \cp \int  D\alpha e^{i (\alpha_{\bar q} +
v \alpha _g) q x} {\cal V}_\perp(\alpha_i)~.
\eea
In Eq. (\ref{e10527}) we have,
\bea
\label{nolabel}
\mu_{\cal P} = f_{\cal P} {m_{\cal P}^2\over m_{q_1} + m_{q_2}}~,~~~~~
\widetilde{\mu}_{\cal P} = {m_{q_1} + m_{q_2} \over m_{\cal P}}~, \nnb
\eea
and $D\alpha = d\alpha_{\bar q} d\alpha_q d\alpha_g
\delta(1-\alpha_{\bar q} - \alpha_q - \alpha_g)$, and
and the DA's $\varphi_{\cal P}(u),$ $\Bbb{A}(u),$ $\Bbb{B}(u),$
$\varphi_P(u),$ $\varphi_\sigma(u),$
${\cal T}(\alpha_i),$ ${\cal A}_\perp(\alpha_i),$ ${\cal A}_\parallel(\alpha_i),$
${\cal V}_\perp(\alpha_i)$ and ${\cal V}_\parallel(\alpha_i)$
are functions of definite twist whose explicit expressions can be found in
\cite{R10511,R10512,R10513}.

\end{document}